\documentclass[11pt]{article}

\def\mytitle#1{\setcounter{equation}{0}
\setcounter{footnote}{0}
\begin{flushleft}\Large\textbf{#1}\end{flushleft}
\vspace{0.25cm}}
\def\myname#1{\leftline{{\large #1}}\vspace{-0.13cm}}
\def\myplace#1#2{\small\begin{flushleft}\textit{#1}\\
\texttt{#2}\end{flushleft}}

\def\myclassification#1{\small\noindent
Pacs no :
       #1\vspace{0.5cm}}
\usepackage{graphicx}
\begin{document}

\mytitle{Change of Density when Chaplygin gas is accreting upon a black hole}

\vskip0.2cm \myname{Ritabrata Biswas\footnote{biswas.ritabrata@gmail.com}} 
\myplace{Department of Mathematics, Jadavpur University,
Kolkata-700 032, India.} {}

\begin{abstract}
In this work I study the density change along the accretion and wind branches when modified Chaplygin gas is taken to be the accreting fluid. I get a high wind density very near to black hole while a low wind density far from the black hole. In modified Chaplygin gas accretion case a local minima in accretion-density is got which can be detected as the CENBOL.
\end{abstract}

\myclassification{98.62.Mw, 95.36.+x, 98.62.Mw, 95.30.Sf}\\\\
The first steps in the field of study about accretion was taken in the five papers of Hoyle, F., Lyttleton, R.A. and Bondi, H. \cite{Hoyle1,Hoyle2,Hoyle3, Bondi1, Bondi2} where they have started to analyze the accretion of intersteller fluids upon stars in the Newtonian gravity as background. Another particular thing about the Bondi accretion was that it was a spherical one. Michel, F. first time has taken the compact object as the Schwarzschild singularity and tried to study the accretion procedure as a general relativistic phenomenon. In his study the idea of critical point for the first time was noticed and critical conditions were found out. First time the approach twoards binary systems, i.e., disc accretion inspite of a spherical one was suggested by Shakura, N. I. and Sunyaev, R. A. \cite{Shakura1}. Novikov, I. D. and Thorne, K. S. and Page, D. N. \cite{Novikov1, Page1} enriched the literature by studying the disc accretion in general relativistic perspective. Das, T and Chakraborti, S.  \cite{Das1} again have shown that as matter is accreting towards a black hole the centrifugal force increases rapidly as compared to the gravitational force. Hence, matter feels increasing centrifugal force around a radius little away from the black hole, then slows down and piles up around the radius. This region refers to CENBOL (CENtrifugal pressure supported BOundary Layer) acting as the effective boundary layer of the black hole system which, like stellar surface, could produce outflowing winds. Therefore, due to the presence of CENBOL, a black hole system, like star, easily exhibits wind of matter along with inflow. However, once matter crosses the CENBOL, the dominant gravitational force solely controls dynamics and matter
falls into the black hole. The nonlinearity arose due to the inclusion of general relativity is simplified by deriving the pseudo Newtonian potential. Mukhopadhyay, B. \cite{Mukhopadhyay2} has derived a form of pseudopotential for rotating black holes. Using such a potential Mukhopadhyay, B. analysed the accretion phenomenon upon black holes where adiabatic gas was the accreting fluid\cite{Mukhopadhyay3}. 

Now-a-days there is no dispute that our universe is going through an accelerated expansion phase. To explain such an universe we have either to change the left hand side,i.e., the geometry part of Einstein equation by modifying Einstein gravity or to change the right hand side's matter part under the Einstein gravity. Such exotic matters which can create negative pressure are been differentiated into two types : dark energy (DE) and dark matter (DM). Many models are been proposed to describe DE. Chaplygin gas model is such a model. Benaoum, H. B. in 2002 has proposed an equation of state (EoS) for Modified Chaplygin Gas(modified Chaplygin gas) as
\begin{equation}\label{1}
p=\alpha \rho-\frac{\beta}{\rho^{n}}
\end{equation}
As the universe is assumed to be filled up by DE, it is very much obvious that the impact of accretion upon black holes would be a matter of interest. In 2004 Babichev, E. et. al. \cite{Babichev1} studided the accretion of dark fluid upon a spherically symmetric non rotating black hole. Now besides comparative small black holes there exists supermassive galactic black holes. Galaxies centering those black holes are mainly rotating. Though the origin of this rotation is quite unknown still. Some assumptions give the responsibility upon primordial turbulance. Consequently, the angular momentum of dark matter halos and eventually the rotation of 
galaxies is thought to be produced by gravitational tidal torque \cite{Barnes1}. It has been suggested that the halos 
obtain their spins through the cumulative acquisition of angular momentum from satellite accretion. DE that has accreted 
on a galaxy would be similarly torqued by such tidal interactions. Although it is difficult to estimate this 
effect without invoking specific models, it is reasonable to expect that some angular momentum might reside in 
the DE halos of galaxies. If such rotating DE were to accrete on a compact object then it would carry a part of 
its angular momentum with it, thus leading to the formation of a DE accretion disc as opposed to Bondi accretion. 

Biswas, R. et. al. recently \cite{Biswas1} have studided both the spherical and disc accretion upon both rotating or nonrotating black holes and analysed the accretion-wind velocity profiles. They have considered steady state accretion of inviscid fluid where specific angular momentum is constant throughout the flow and got the radial energy momentum conservation equation as  
\begin{equation}\label{2}
u\frac{du}{dx}+\frac{1}{\rho}\frac{dp}{dx}-\frac{\lambda^{2}}{x^{3}}+F_{g}(x)=0,
\end{equation}
where all variables are expressed in dimensionless units as follows:\\
$u=u_{r}=\frac{v}{c},~x=\frac{r}{r_{g}},~r_{g}=\frac{GM}{c^{2}}$, where $c_{s}$, $M$ and $c$ are the speed of sound, mass of the black hole and the speed of light respectively and $r$, $v$ are dimensionful radial coordinate and radial velocity respectively.
Here $p$ and $\rho$ are the dimensionless isotropic pressure and density of the flow, $F_{g}=\frac{\left(x^{2}-2 j\sqrt{x}+j^{2}\right)^{2}}{x^{3}\left(\sqrt{x}(x-2)+j\right)^{2}}$, is the gravitational force corresponding 
to the pseudo-Newtonian potential  \cite{Mukhopadhyay2} and $\lambda$ is the dimensionless specific angular momentum 
(in units of $\frac{GM}{c}$) of the flow, $j$ is the dimensionless specific angular momentum of the black hole (Kerr parameter). In the
case of Bondi flow $\lambda=0$ throughout. 


The equation of continuity is given by
\begin{equation}\label{3}
\frac{d}{dx}\left(x u \Sigma\right)=0,
\end{equation}
where
$\Sigma$, the vertically integrated density, is given by \cite{Santos}
$
\Sigma~~=~~I_{c}\rho_{e}h(x),
$
when,
$I_{c}$  =  constant (related to  the  equation of state of  the accreting fluid) \cite{Mukhopadhyay3},
$\rho_{e}$ = density  at  equatorial  plane and 
$h(x)$  = half-thickness  of  the  disc.
Assuming the vertical equilibrium, the expression for $h(x)$ can be written as:
$
h(x)=c_{s}\sqrt{\frac{x}{F_{g}}}
$
where $c_{s}^{2}= \frac{\partial p}{\partial \rho}\sim\frac{p}{\rho}$ is the square of sound speed. 

Integrating (\ref{4}) we have
\begin{equation}\label{4}
\dot{M}=\odot \rho c_{s} \frac{x^{\frac{3}{2}}}{F_{g}^{\frac{1}{2}}}u=\odot c_{s}u\left(\frac{x^{3}}{F_{g}}\right)^{\frac{1}{2}}\left(\frac{n \beta}{c_{s}^{2}-\alpha}\right)^{\left(\frac{1}{n+1}\right)},
\end{equation}
where $\odot$ is a geometric constant, depending on the exact geometry of the flow. Taking log and differentiating the equation for sound-speed-square's gradiant becomes
\begin{equation}\label{5}
\frac{\left(1-n\right)c_{s}^{2}+\alpha\left(n+1\right)}{\left(n+1\right)c_{s}\left(c_{s}^{2}-\alpha\right)}\frac{dc_{s}}{dx}=\frac{3}{2x}-\frac{1}{2F}\frac{dF}{dx}+\frac{1}{u}\frac{du}{dx}
\end{equation}

From the EoS the pressure gradiant can be calculated as
\begin{equation}\label{6}
\frac{1}{\rho}\frac{dp}{dx}=-\frac{1}{n+1}\frac{d}{dx}\left(c_{s}^{2}\right)-\left(\frac{\alpha}{n+1}\right)\frac{d}{dx}\left\{\ln\left(c_{s}^{2}-\alpha\right)\right\}.
\end{equation}
replacing in (\ref{2}) and with the help of (\ref{6}) the readial velocity gradiant becomes
\begin{equation}\label{7}
\frac{du}{dx}=\frac{\frac{\lambda^{2}}{x^{3}}-F_{g}(x)+\left(\frac{3}{x}-\frac{1}{F_{g}}\frac{dF_{g}}{dx}\right)\frac{c_{s}^{4}}{\left\{\left(1-n\right)c_{s}^{2}+\alpha\left(n+1\right)\right\}}}{u-\frac{2c_{s}^{4}}{u\left\{\left(1-n\right)c_{s}^{2}+\alpha\left(n+1\right)\right\}}}=\frac{N}{D}.
\end{equation}
Equation (\ref{7}) has two branches of solution, accretion (inflow) and wind (outflow). As we are considering a transonic flow so there will be a point where both $N=D=0$ \footnote{Giving the expression for critical point velocity or sound speed immediately as a solution of two different algebraic equations.}. Using L'Hospital's rule we can obtain the velocity gradiant at that critical point. 
In the reference \cite{Biswas1} the physically important velocity profiles are of adibatic and Chaplygin gas flow of Bondi type, disc flow upon  Schwarzschild black hole and Kerr type black hole with rotational parameters $j=0.5~and ~0.9$ respectively. Here I have given these curves in the figures 1(a) to 1(h).  

From different literature we can collect the observational data supported values of $\alpha$ and $n$ as
\begin{center}
{\bf Table I:} Permissible ranges of modified Chaplygin gas parameters in different models.\\
\begin{tabular}{|l|}
\hline\hline
~~Source of data
~~~~~~~~~~~~~~~~~~~~~~~~~~~~~~~~~~~~~~~~~~~~~~~~~~~~~~~~Type of Model \\ \hline
\\
~~~~~~~~~~~~~~~~~~~~~~~~~~~~~~~~~~~~~~~~~~~~~~~~~~~~~~~~~~~~~~CDM~~~~~~~~~~~~~~~~~~~~~~~~~~~~~~~~UDME~~~~~~~
\\\hline
~~~~~~~~~~~~~~~~~~~~~~~~~~~~~~~~~~~~~~~~~~~~~~~~~~~~~~$\alpha$~~~~~~~~~~~~~~~~~~$n$~~~~~~~~~~~~~~~~~~~~$\alpha$~~~~~~~~~~~~~~~~~~$n$~~~~~~~
\\\hline\hline
\\
~$H(z)-z$~~~~~~~~~~~~~~~~~~~~~~~~~~~~~~~~~$0\leq\alpha\leq 1.07$~~~~~~$0\leq n\leq 1$~~~~~~~$0\leq\alpha\leq 1.35$~~~~~$0\leq n\leq 1$
\\\\\hline\\
~$H_{0}-t_{0}$~~~~~~~~~~~~~~~~~~~~~~~~~~~~~~~~~~~~~~$\alpha=0.01$~~~~~~~~~$ n=0.01$~~~~~~~~~~$\alpha=0.06$~~~~~~~~~$ n=0.11$~\\
(best-fit)
\\\\ \hline\hline
\end{tabular}
\end{center}
In 1(i)-(l) the velocity profile for best fit parameter range from the above table.
\begin{figure}
~~~~~~~~~~(a)~~~~~~~~~~~~~~~~~~~~~~~~~~~~~(b)~~~~~~~~~~~~~~~~~~~~~~~~~~~~~(c)~~~~~~~~~~~~~~~~~~~~~~~~~~~~~(d) \\
\includegraphics[height=1.6in, width=1.6in]{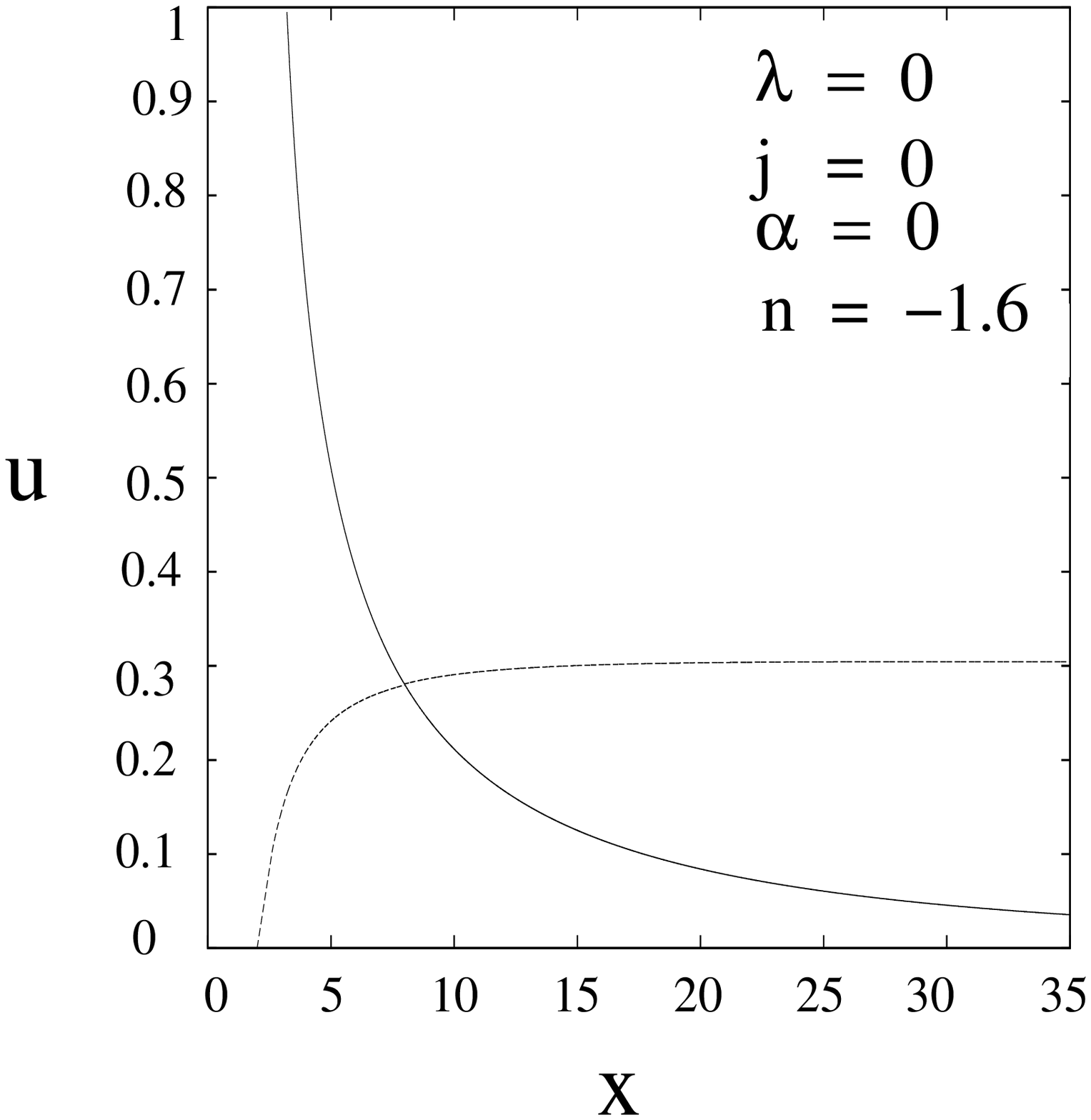}~~\includegraphics[height=1.6in, width=1.6in]{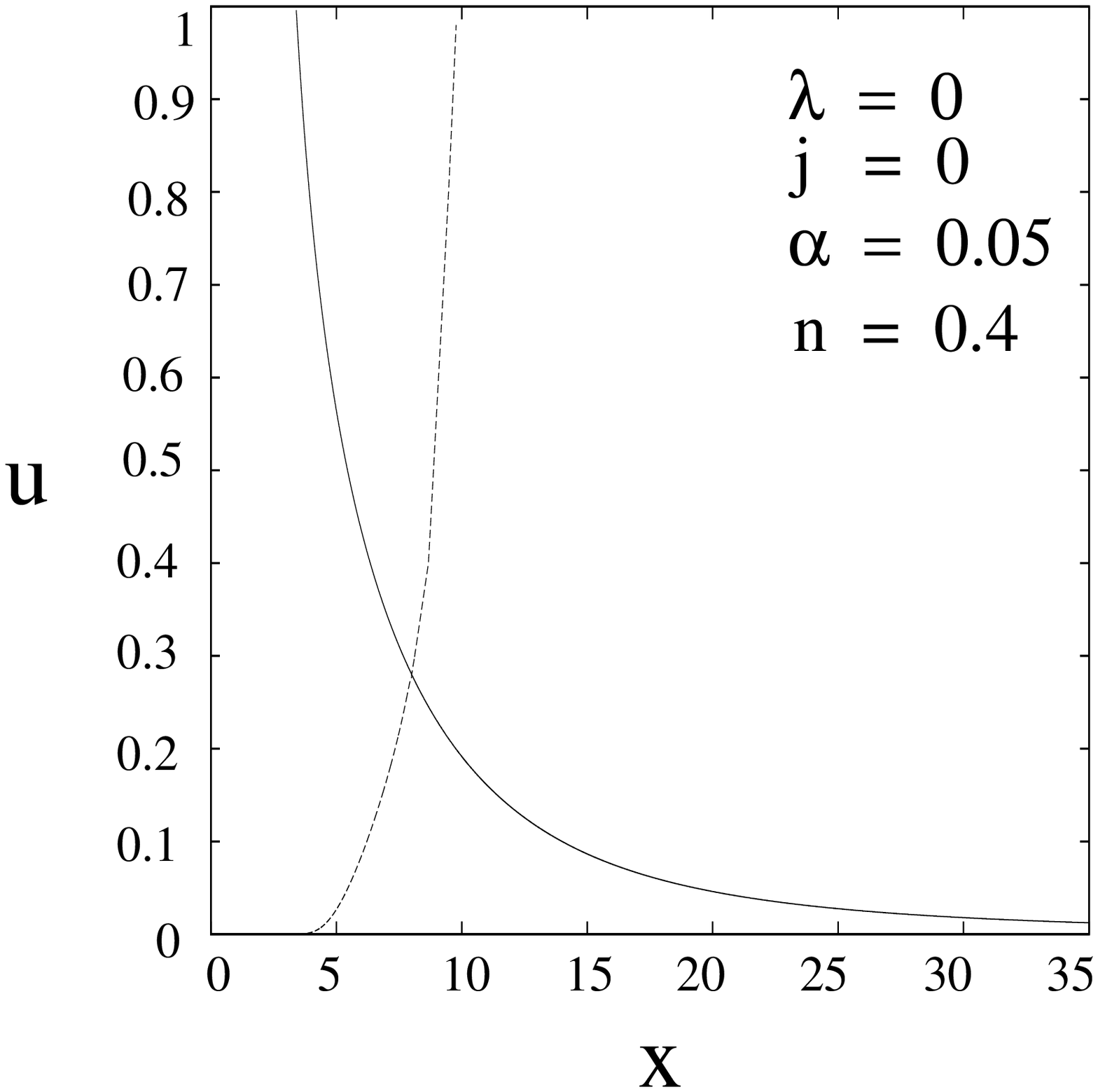}
\includegraphics[height=1.6in, width=1.6in]{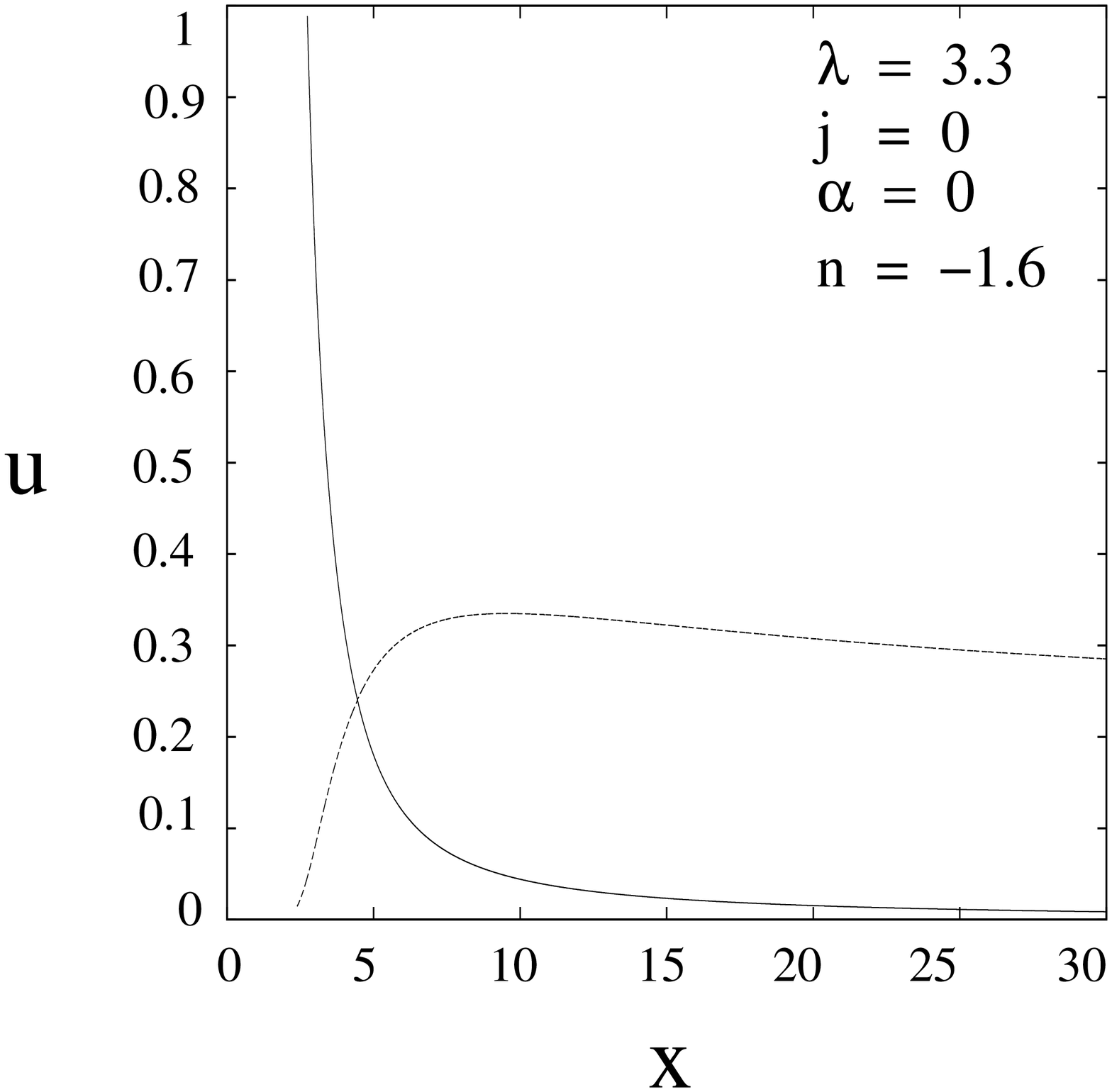}~~\includegraphics[height=1.6in, width=1.6in]{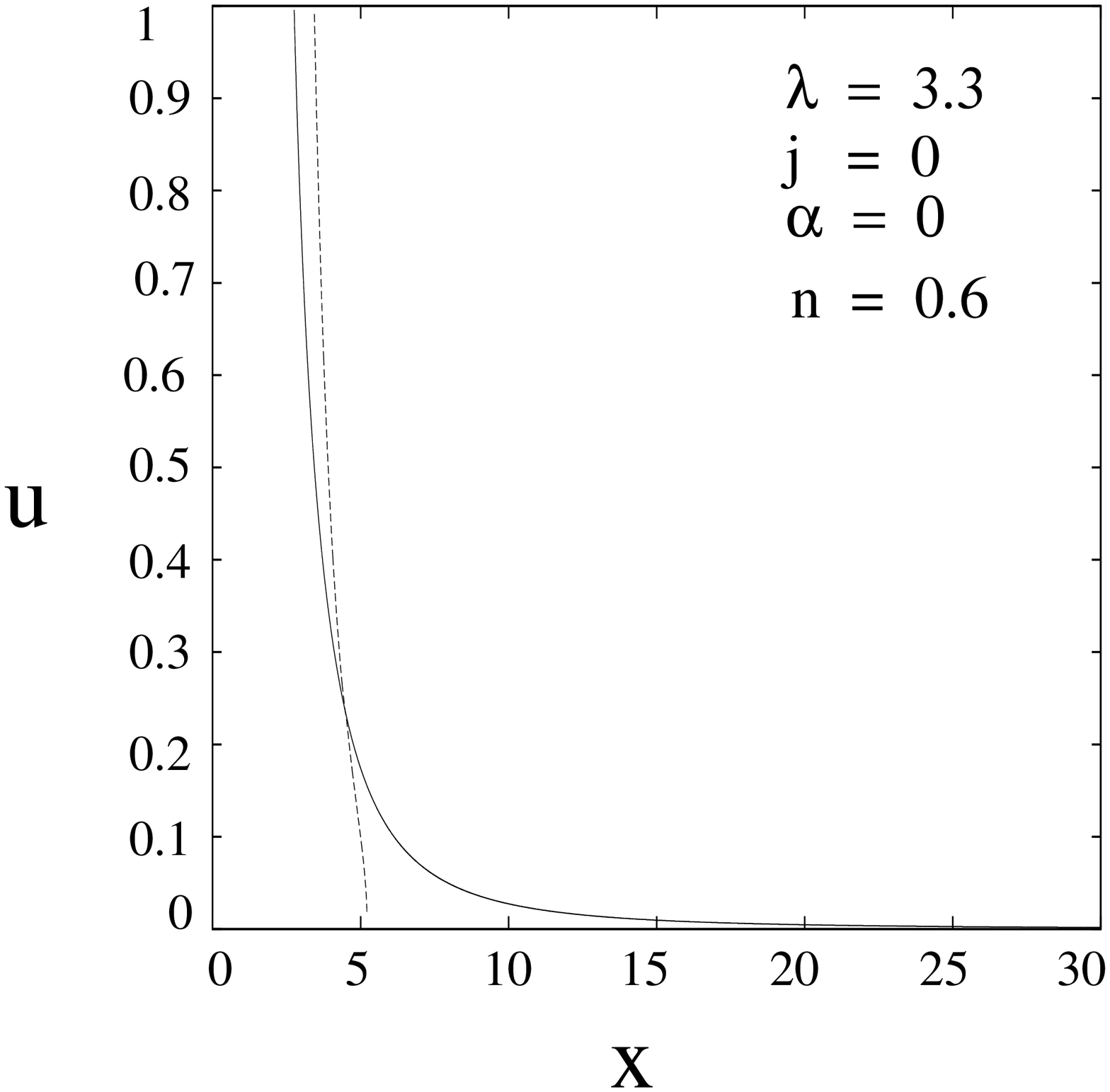}\\
\hspace{1cm}
~~~~~~~~~~(e)~~~~~~~~~~~~~~~~~~~~~~~~~~~~~(f)~~~~~~~~~~~~~~~~~~~~~~~~~~~~~(g)~~~~~~~~~~~~~~~~~~~~~~~~~~~~~(h)\\
\includegraphics[height=1.6in, width=1.6in]{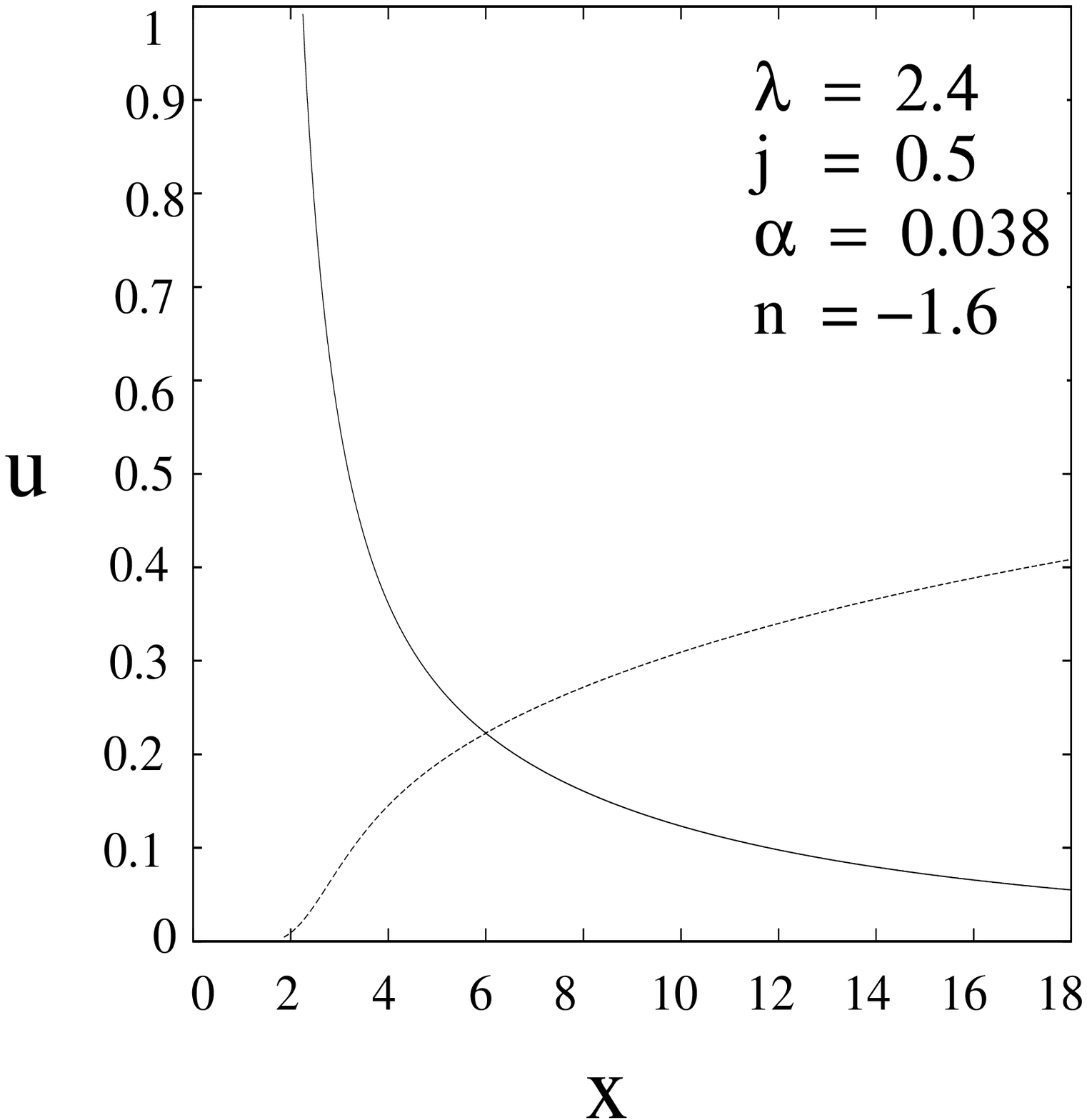}~~\includegraphics[height=1.6in, width=1.6in]{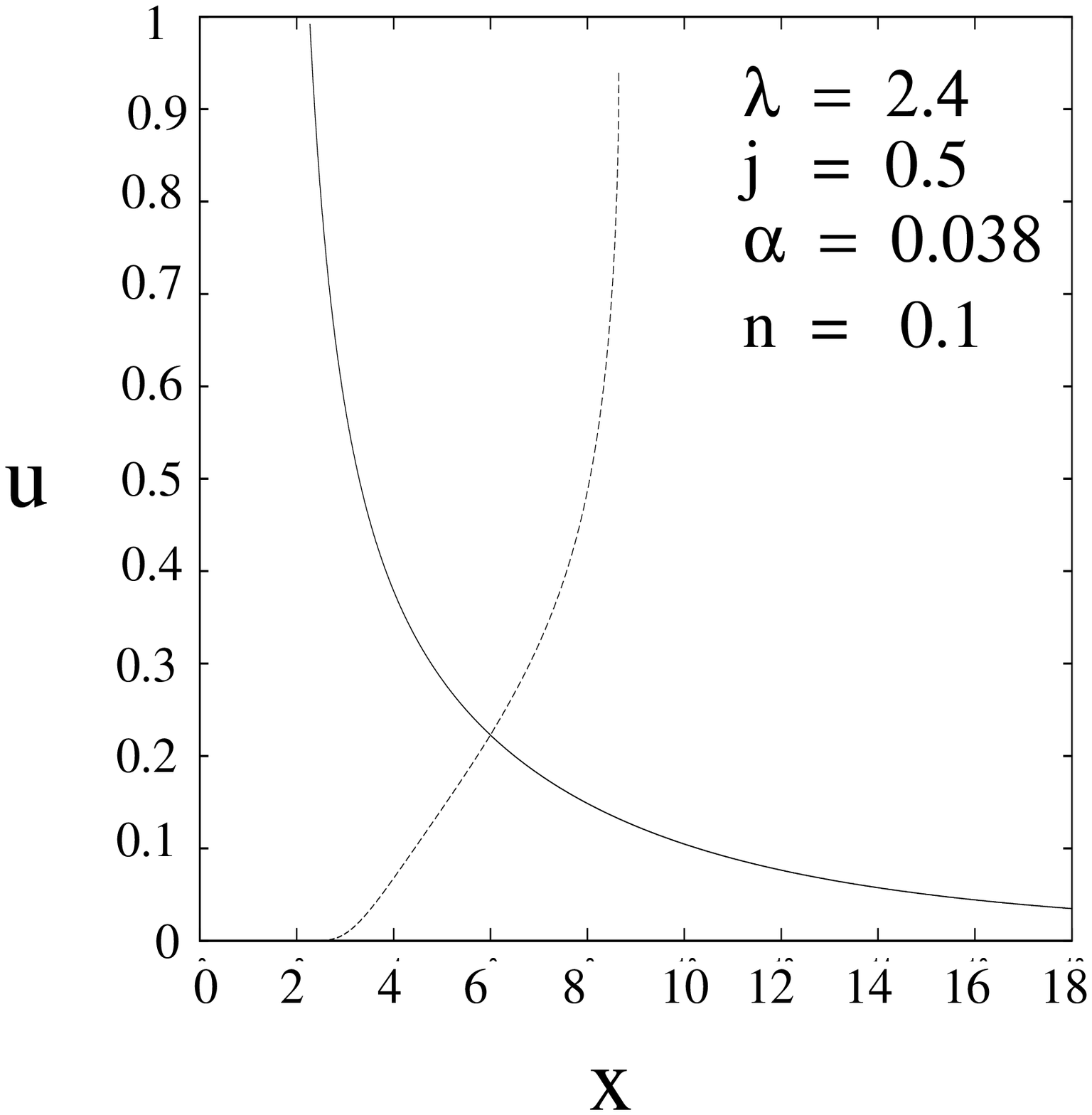}
\includegraphics[height=1.6in, width=1.6in]{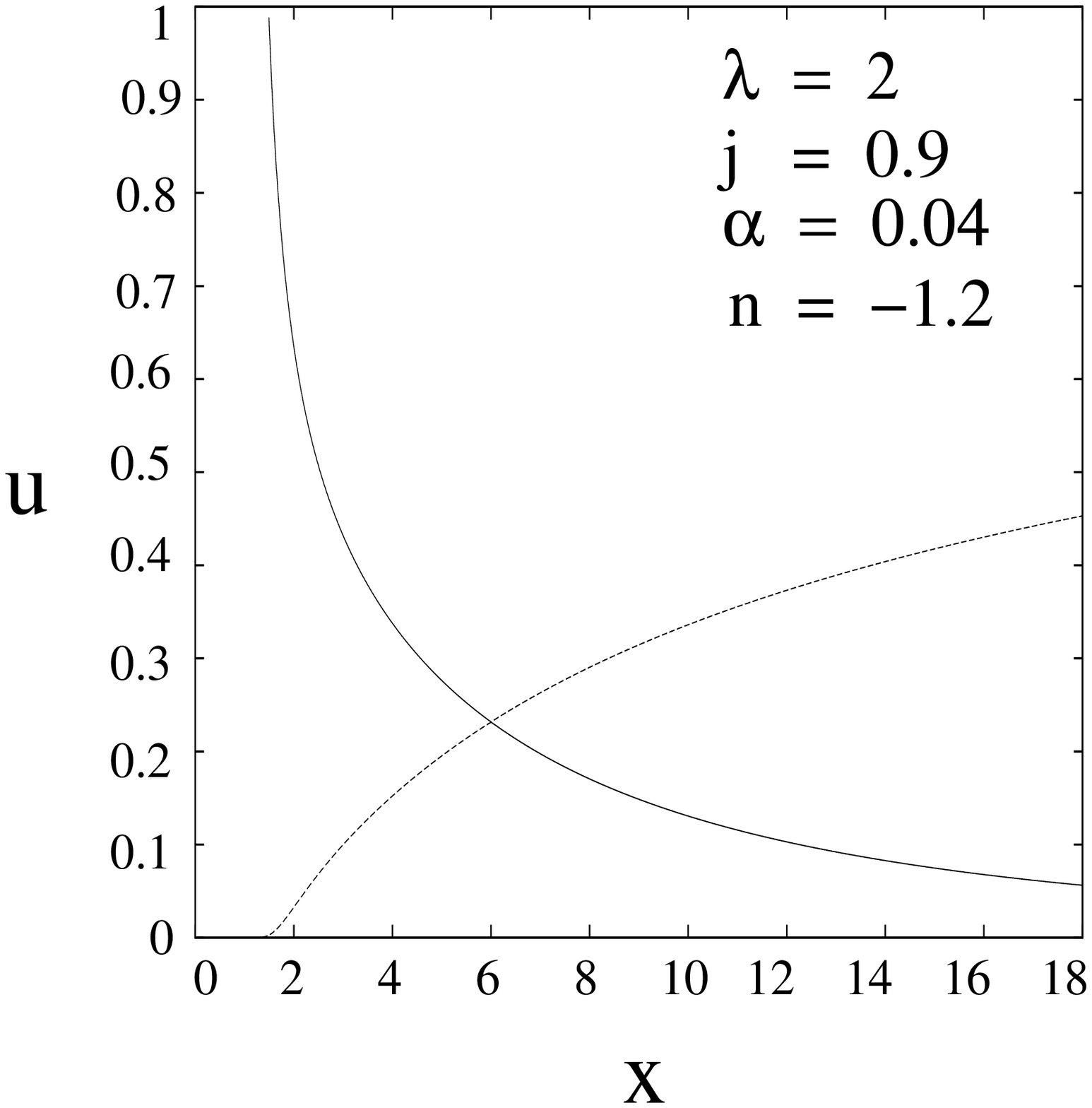}~~\includegraphics[height=1.6in, width=1.6in]{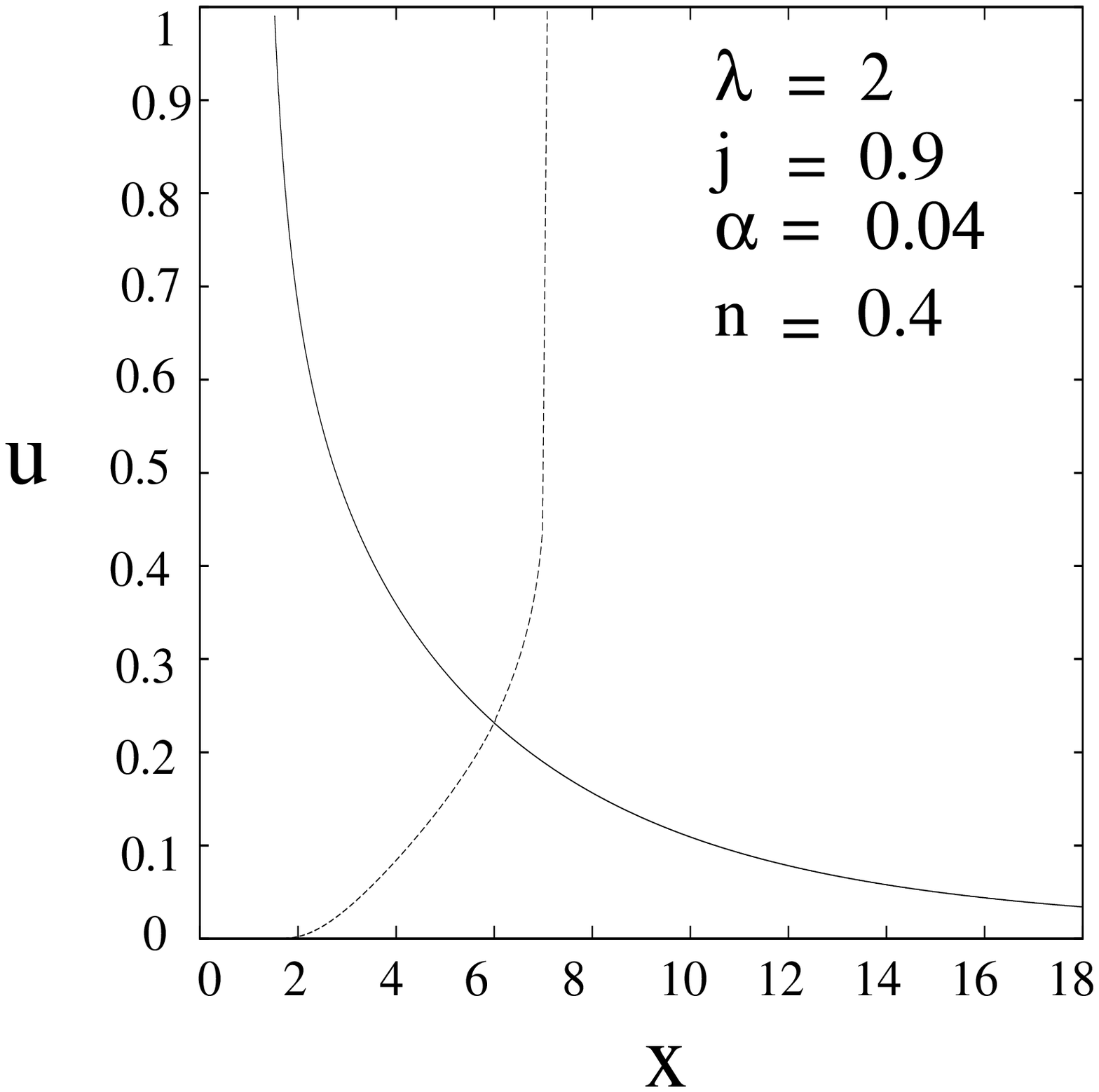}\\
\hspace{1cm}
~~~~~~~~~~(i)~~~~~~~~~~~~~~~~~~~~~~~~~~~~~(j)~~~~~~~~~~~~~~~~~~~~~~~~~~~~~(k)~~~~~~~~~~~~~~~~~~~~~~~~~~~~~(l)\\
\includegraphics[height=1.6in, width=1.6in]{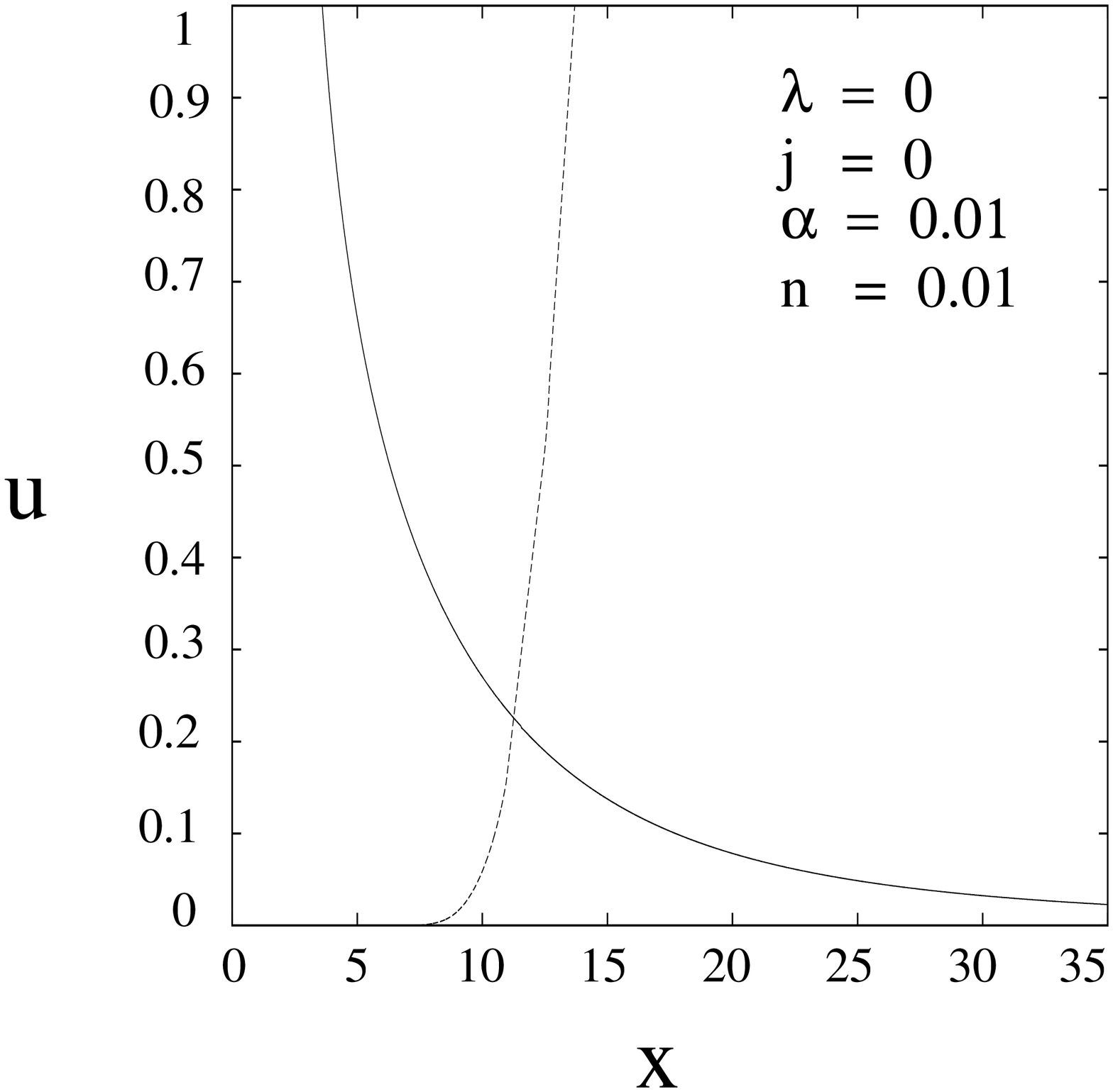}~~\includegraphics[height=1.6in, width=1.6in]{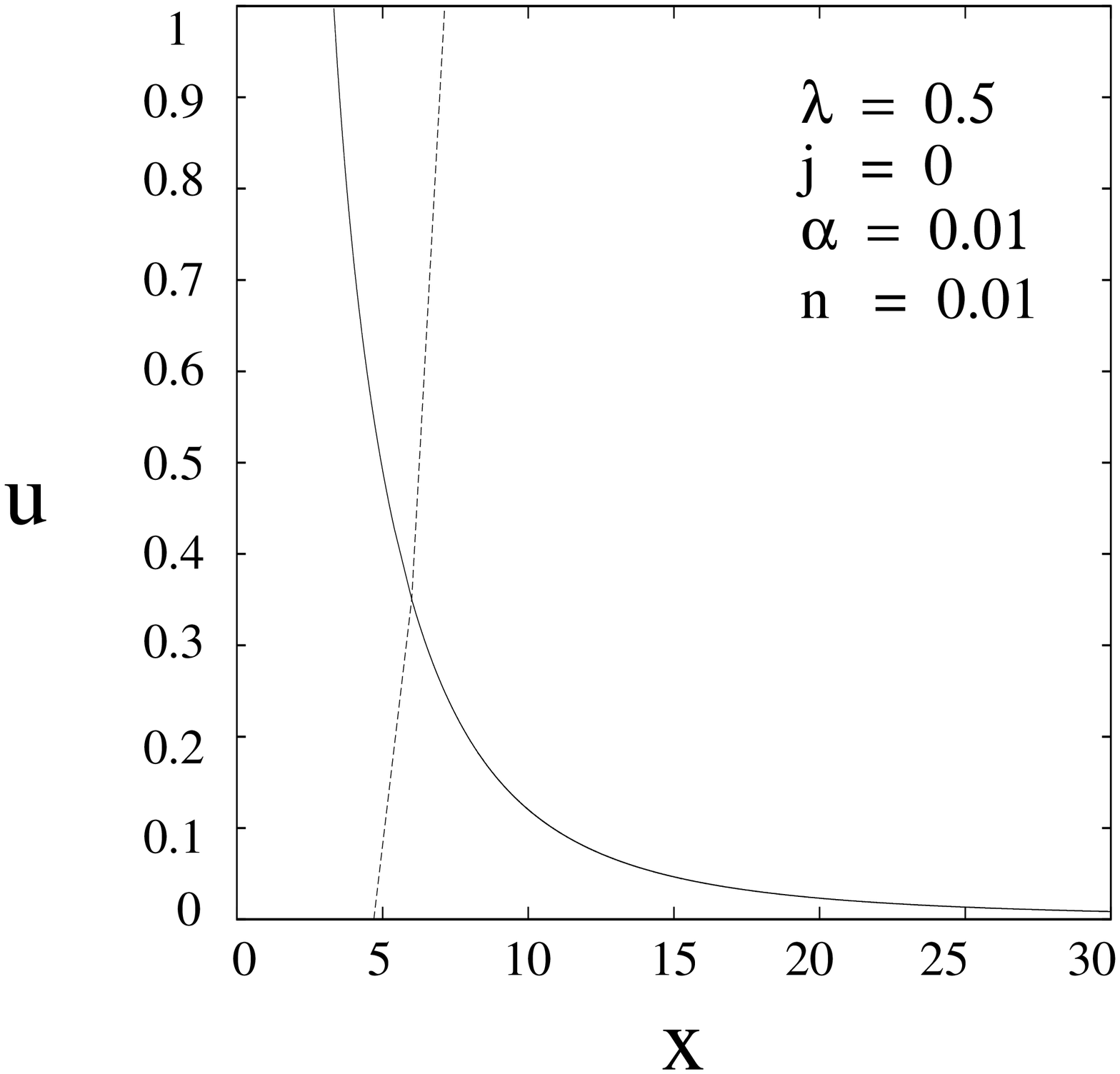}
\includegraphics[height=1.6in, width=1.6in]{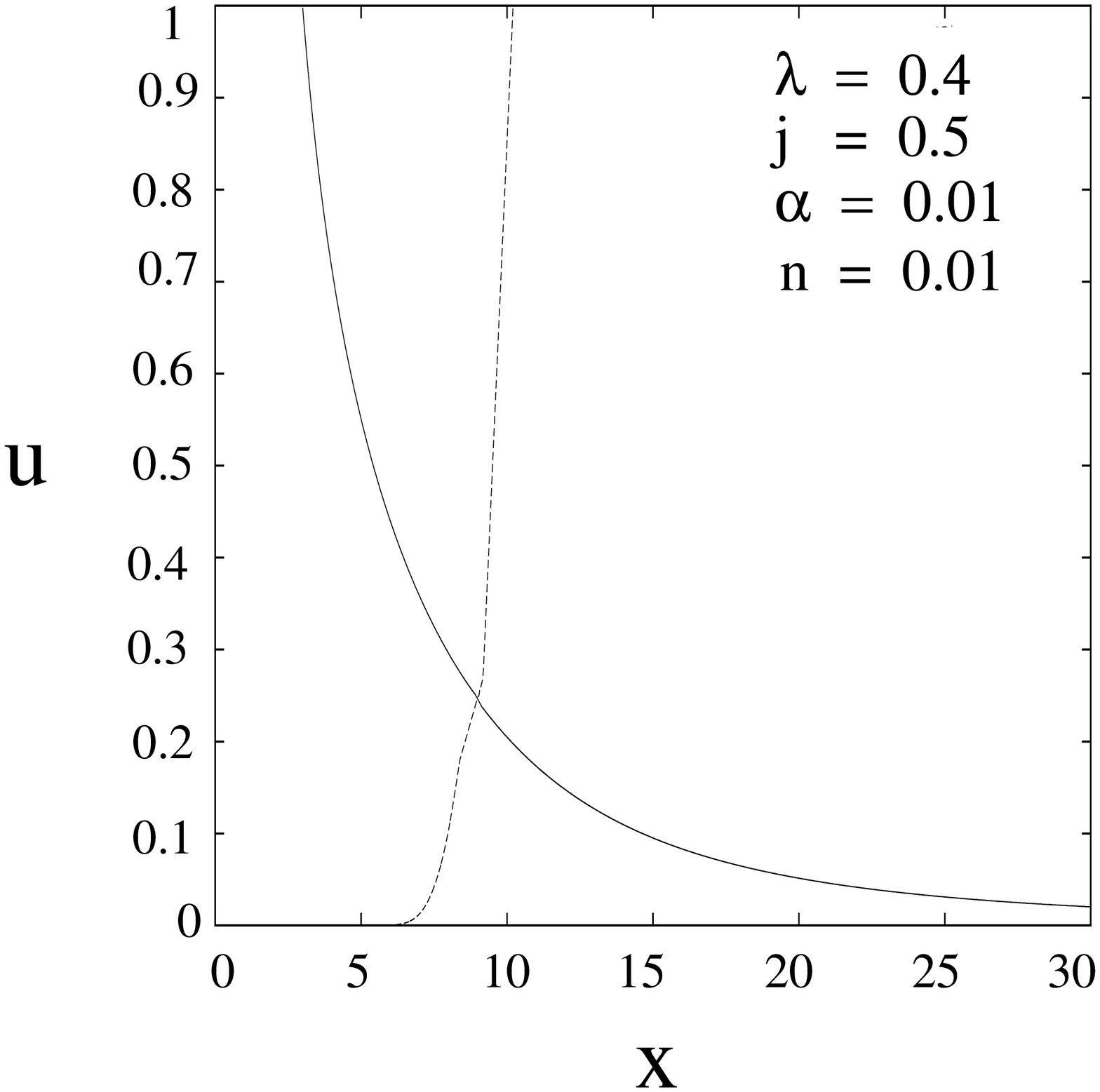}~~\includegraphics[height=1.6in, width=1.6in]{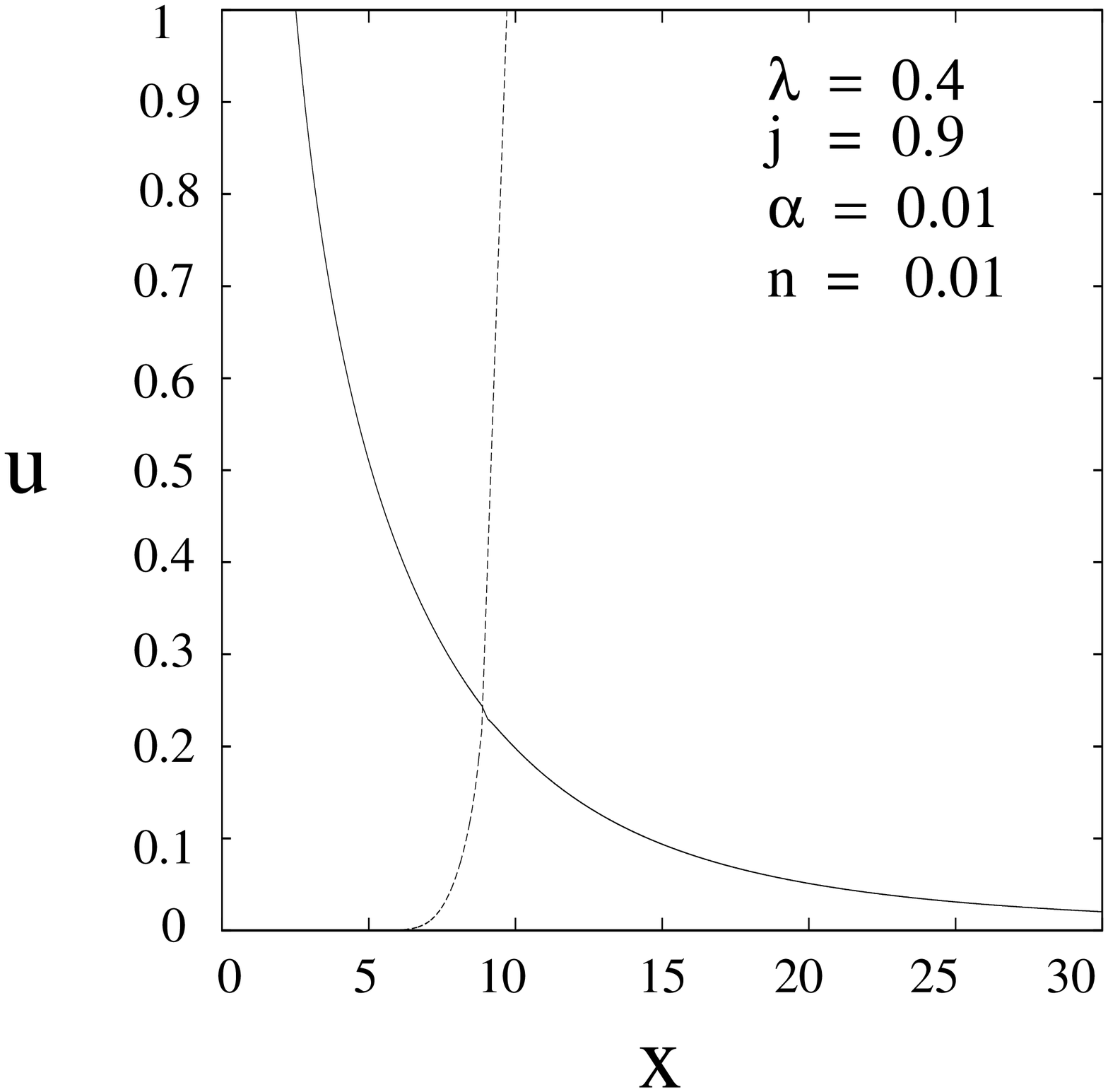}\\
\hspace{1cm}
Fig. 1(a)-1(l) represent the velocity profiles for different parametrical values. The solid lines represent the accretion whereas the dotted lines are for wind.\hspace{1cm} \vspace{2cm}

\end{figure}\\\\
We know for spherical accretion
\begin{equation}\label{15th}
\dot{M}=4\pi \rho v r^{2}
\end{equation}
where $\dot{M}$ is the change in black hole mass. and the Edington mass
accretion rate tells us that
\begin{equation}\label{16th}
\dot{M}=\frac{10^{17}M}{M_{\odot}}
\end{equation}
$\dot{M}$ is the solar mass= $1.989\times10^{33}$ gm.

Now using these two equations if we want to calculate the original fluid density in the accretion flow then we must take the values of the units chosen in account.
So we get $$r=x.\frac{GM}{c^{2}} ~and ~ v=u.c$$
So we get the expression of $\rho$ as
\begin{equation}\label{16th}
\rho=\frac{10^{17}c^{3}}{4\pi M_{\odot}G^{2}}.\frac{1}{M}.\frac{1}{u}.\frac{1}{x^{2}}
\end{equation}
Now, as we know $$~~G = 6.67 \times 10^{-8} gm^{-1} cm^{3} s^{-2}~~, ~~c=3\times 10^{10} cm s^{-2}$$
and letting $M=10M_{\odot}$ ,i.e.,  taking the mass of the black hole as ten times of the solar mass we get,
\begin{equation}\label{16th}
\rho=1.2214\times 10^{-6}\times \frac{1}{ux^{2}}
\end{equation}

The density-accretion rate equation for disc acretion is given by (\ref{4}) which with the help of (\ref{16th}) becomes 
\begin{equation}
\rho=\frac{\odot}{10^{18}}\frac{\left(xF_{g}\right)^{\frac{1}{2}}}{c_{s}}\frac{1}{ux^{2}}
\end{equation}

\begin{figure}
~~~~~~~~~~(a)~~~~~~~~~~~~~~~~~~~~~~~~~~~~~(b)~~~~~~~~~~~~~~~~~~~~~~~~~~~~~(c)~~~~~~~~~~~~~~~~~~~~~~~~~~~~~(d) \\
\includegraphics[height=1.6in, width=1.6in]{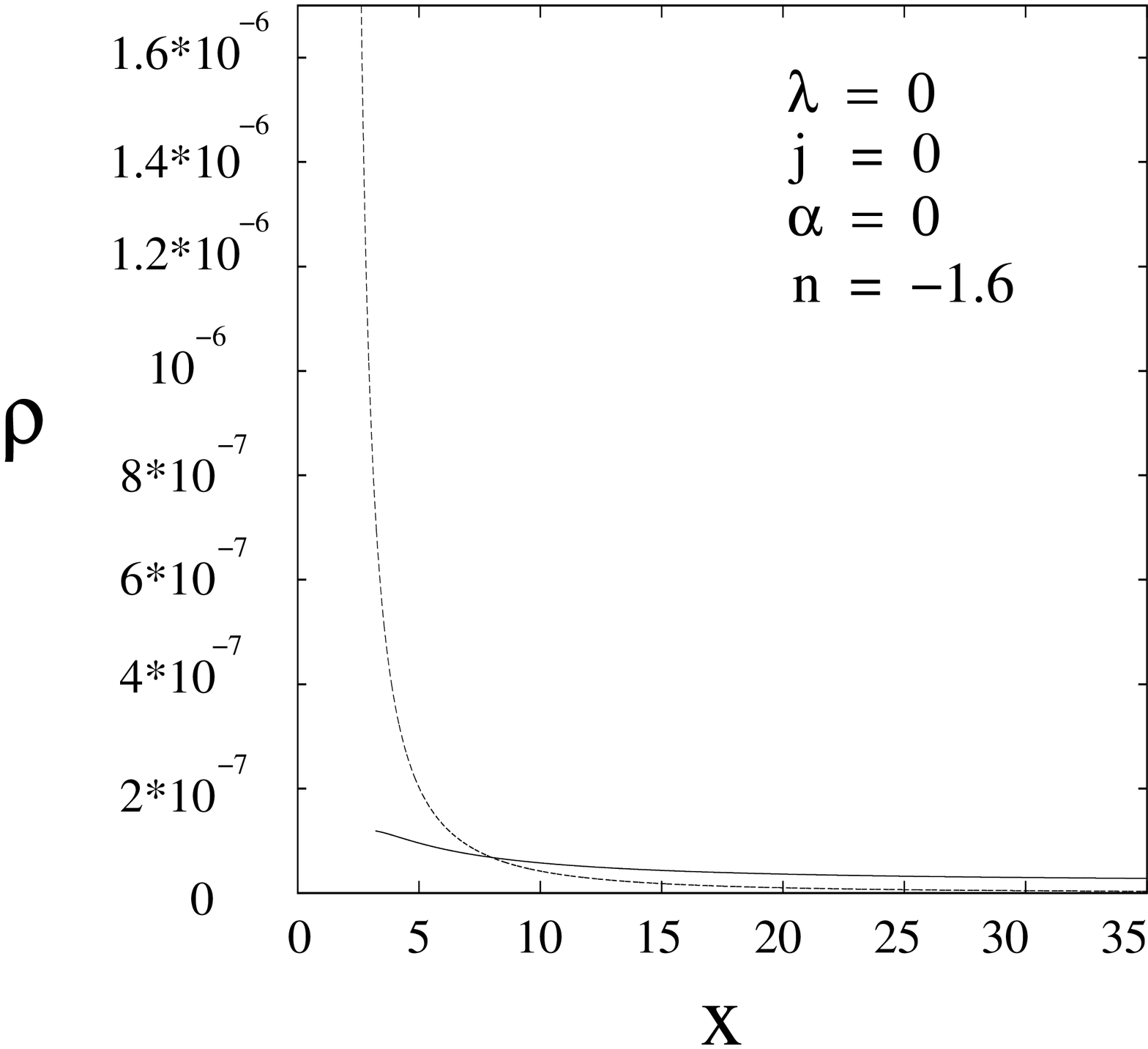}~~\includegraphics[height=1.6in, width=1.6in]{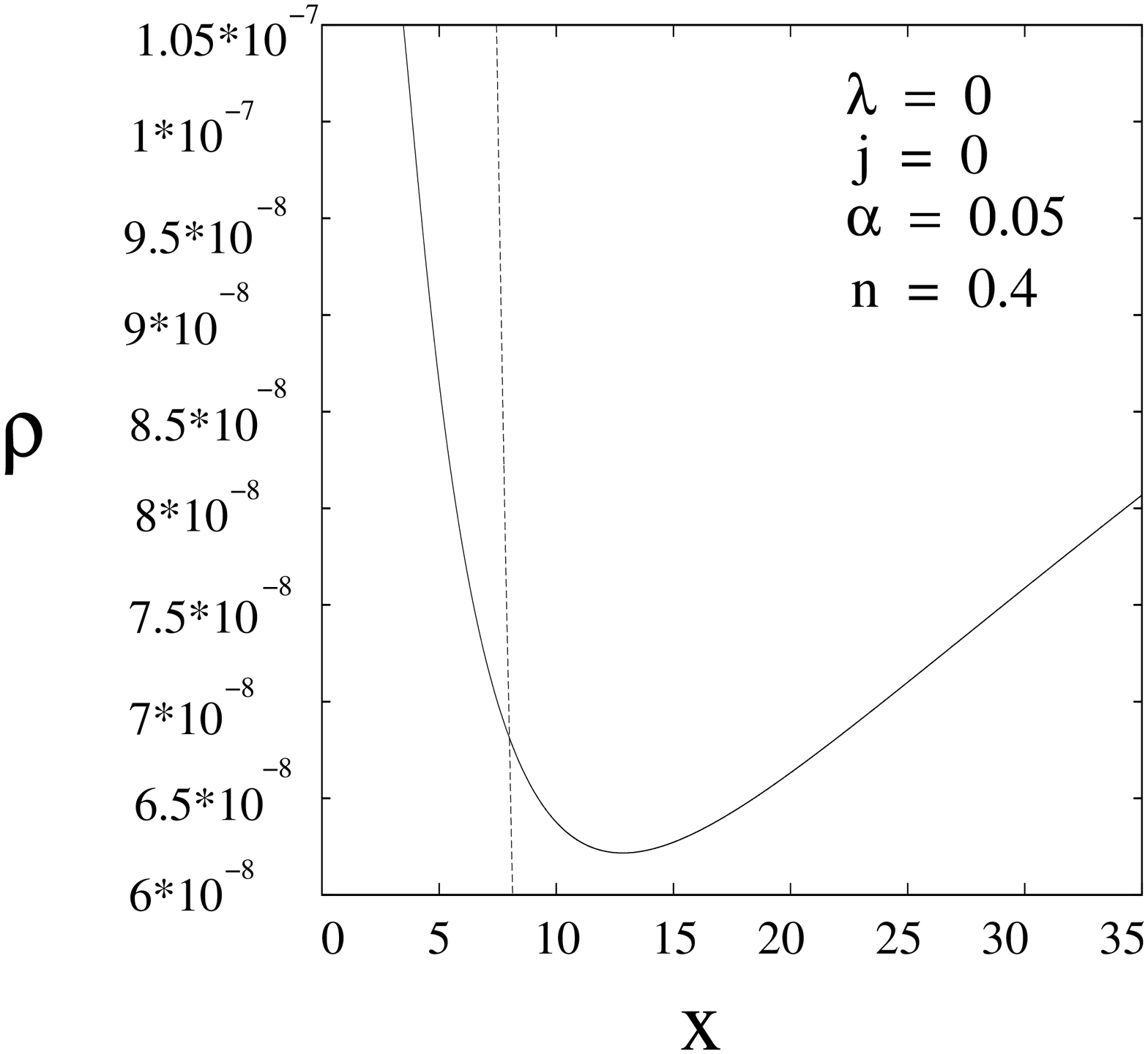}
\includegraphics[height=1.6in, width=1.6in]{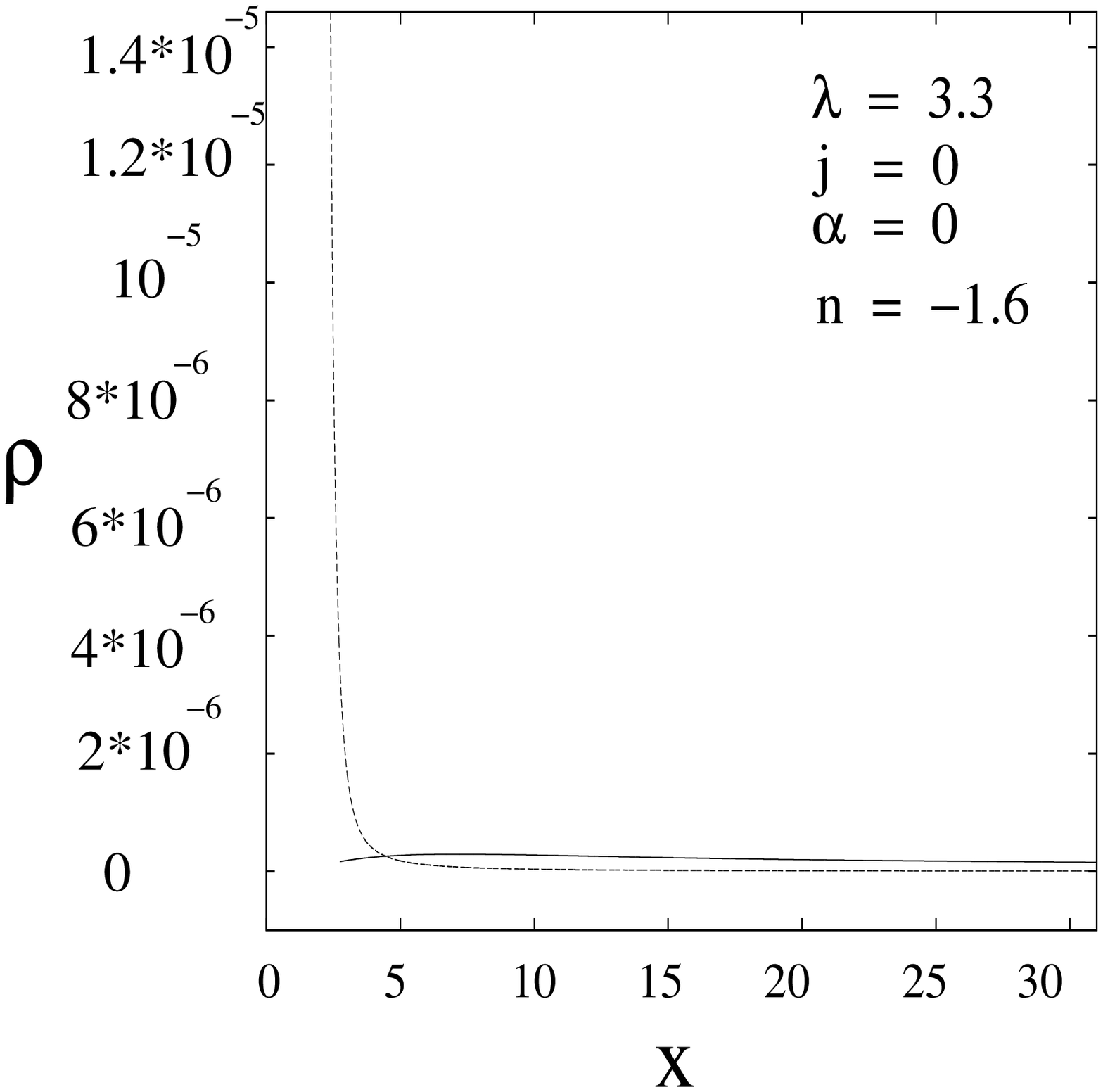}~~\includegraphics[height=1.6in, width=1.6in]{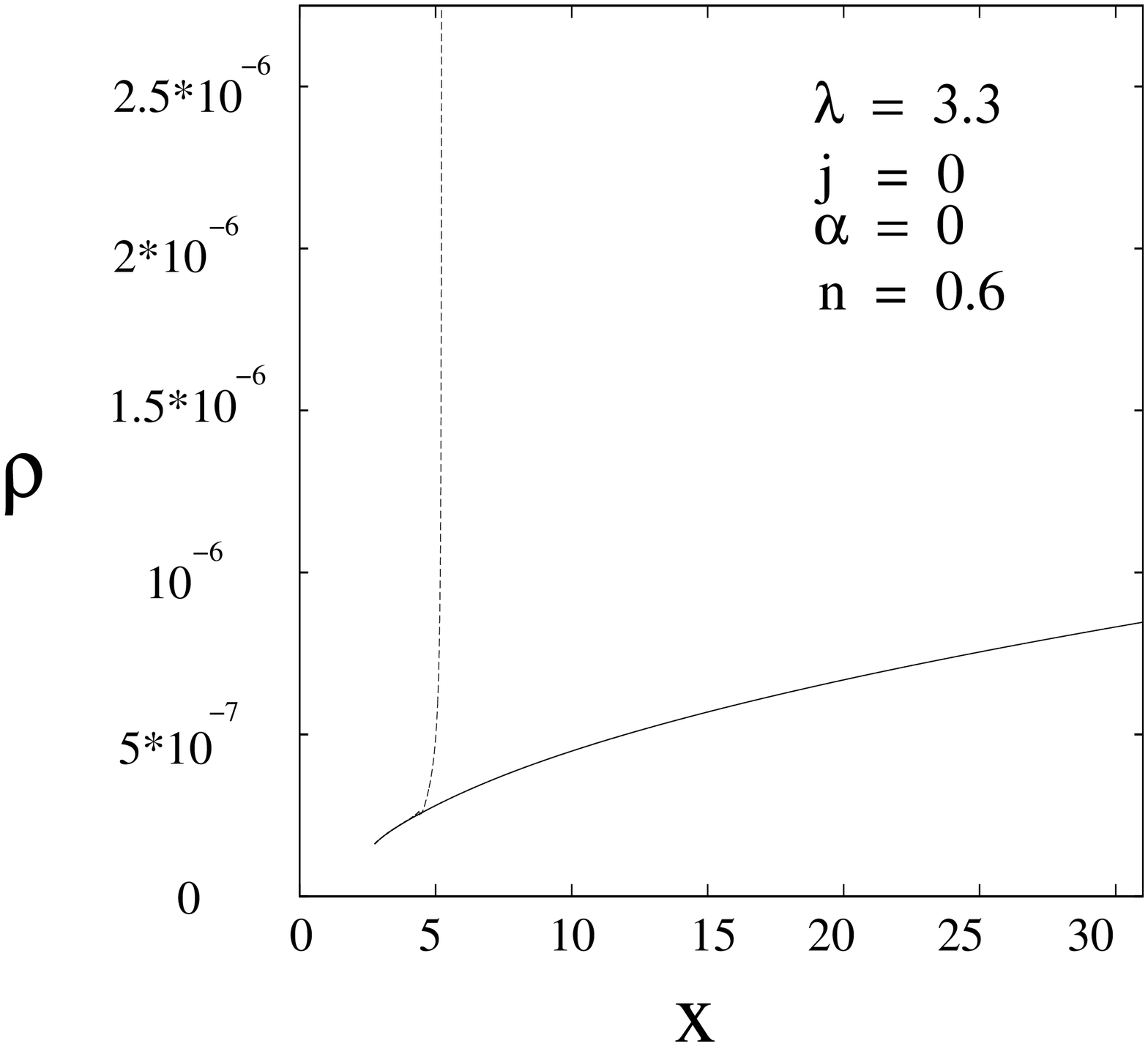}\\
\hspace{1cm}
~~~~~~~~~~(e)~~~~~~~~~~~~~~~~~~~~~~~~~~~~~(f)~~~~~~~~~~~~~~~~~~~~~~~~~~~~~(g)~~~~~~~~~~~~~~~~~~~~~~~~~~~~~(h)\\
\includegraphics[height=1.6in, width=1.6in]{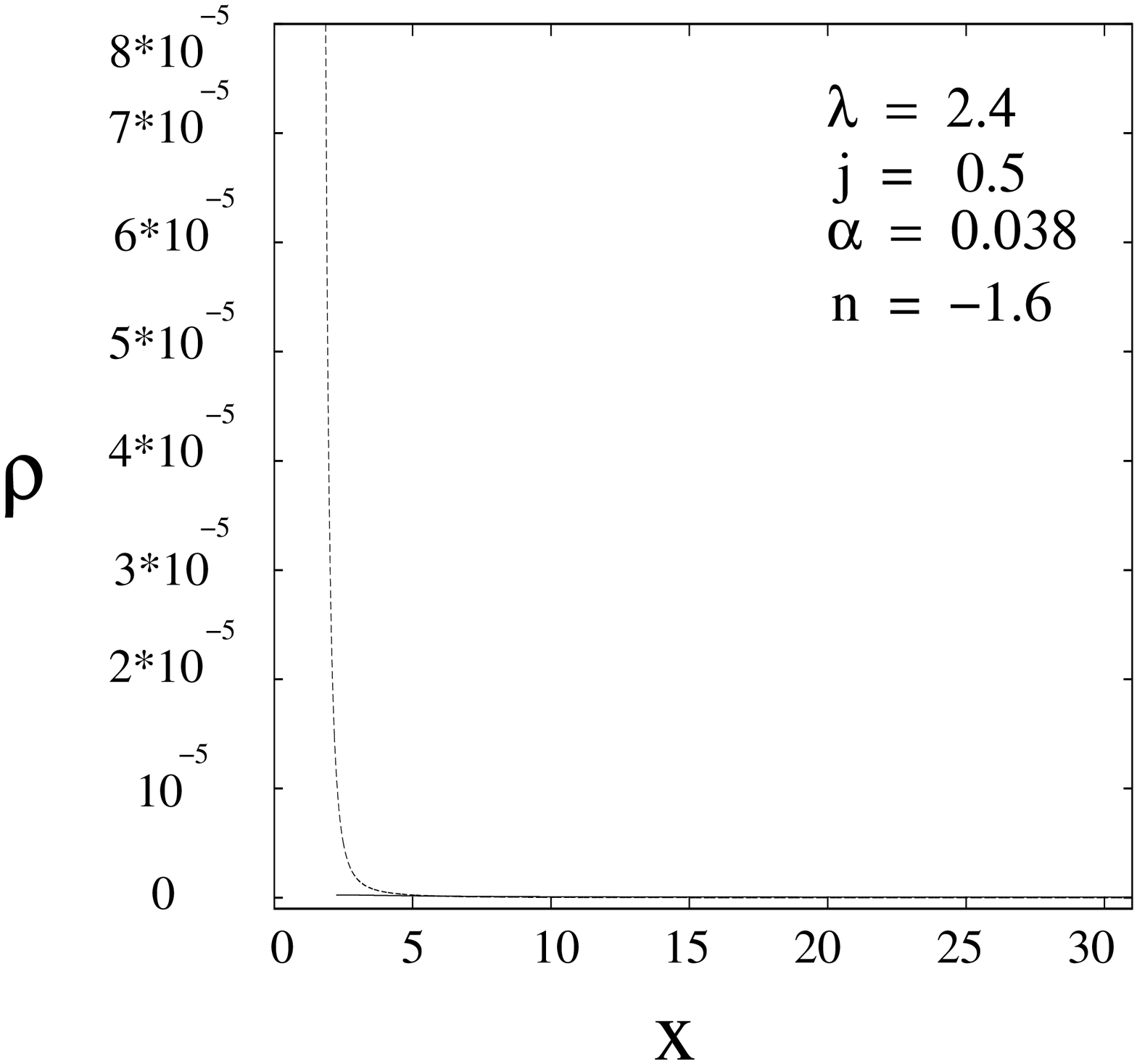}~~\includegraphics[height=1.6in, width=1.6in]{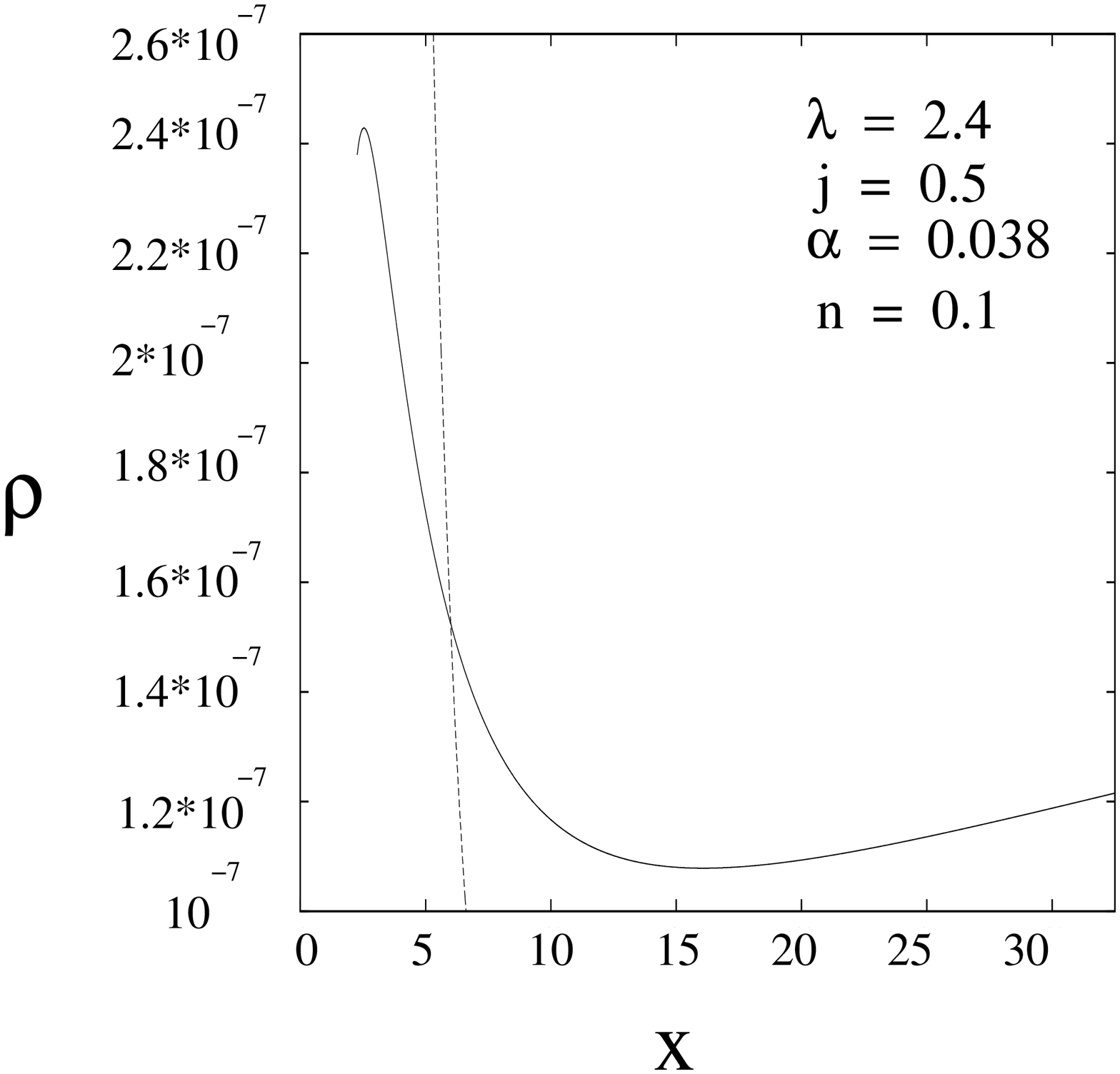}
\includegraphics[height=1.6in, width=1.6in]{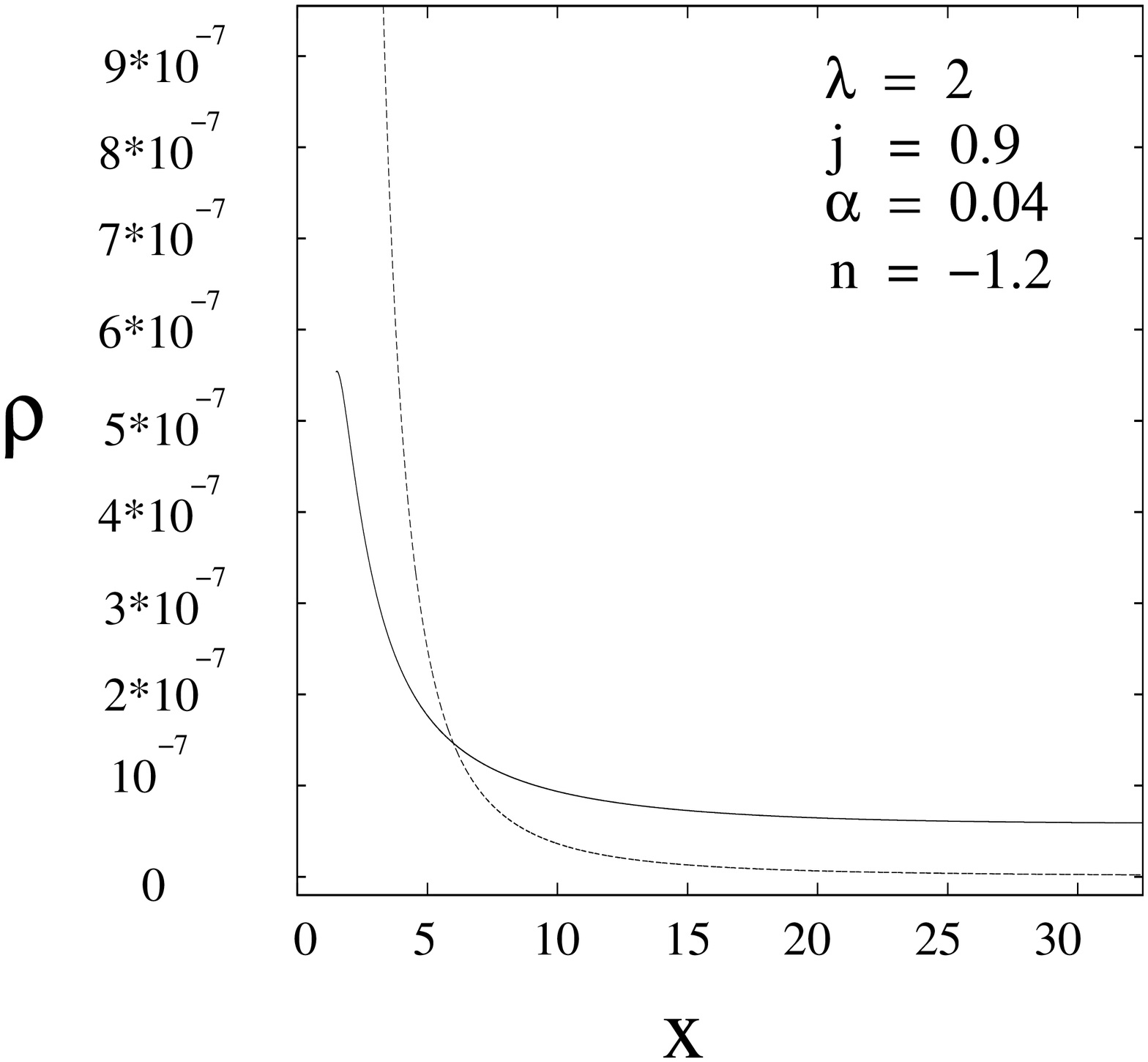}~~\includegraphics[height=1.6in, width=1.6in]{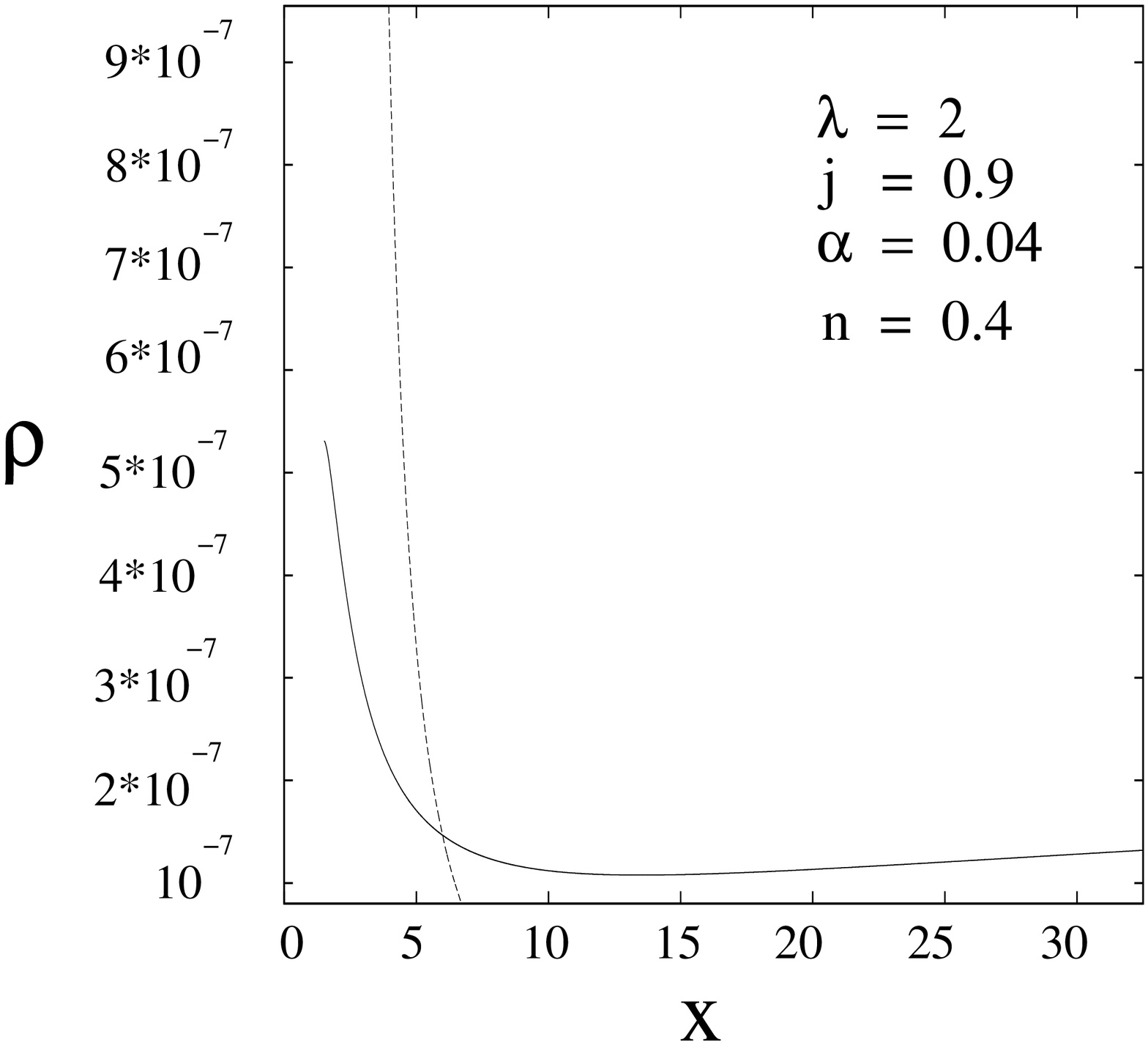}\\
\hspace{1cm}
~~~~~~~~~~(i)~~~~~~~~~~~~~~~~~~~~~~~~~~~~~(j)~~~~~~~~~~~~~~~~~~~~~~~~~~~~~(k)~~~~~~~~~~~~~~~~~~~~~~~~~~~~~(l)\\
\includegraphics[height=1.6in, width=1.6in]{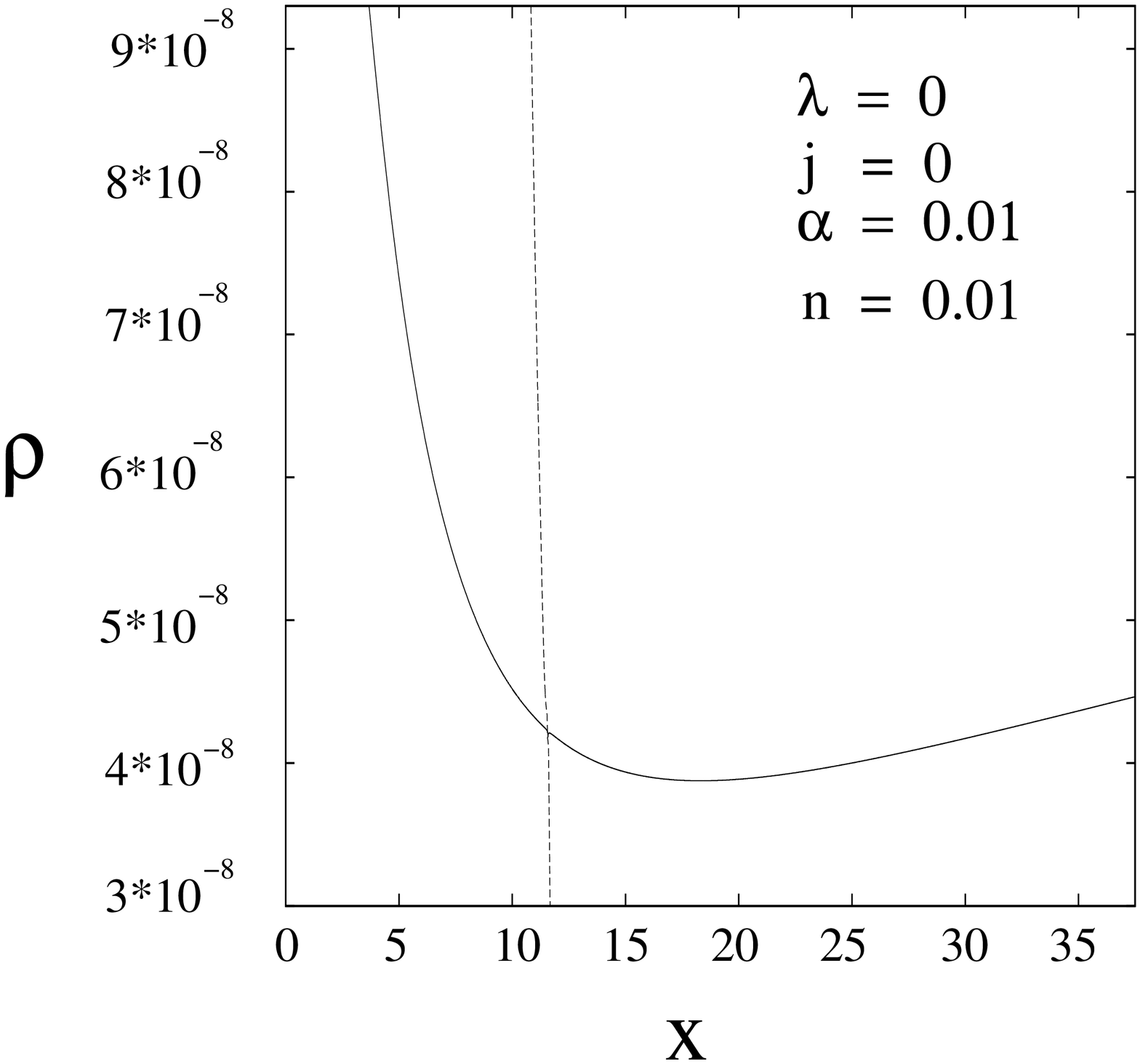}~~\includegraphics[height=1.6in, width=1.6in]{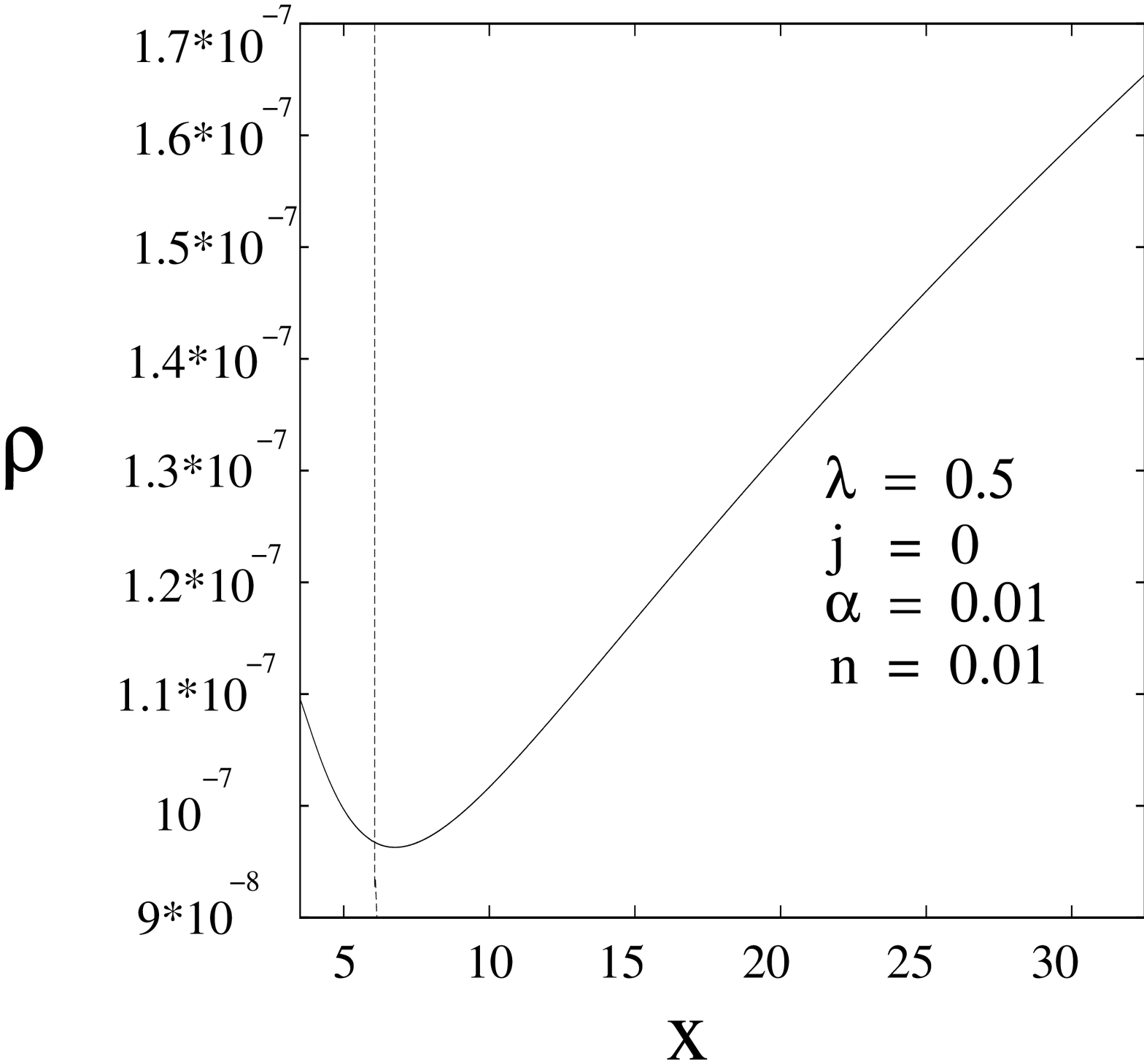}
\includegraphics[height=1.6in, width=1.6in]{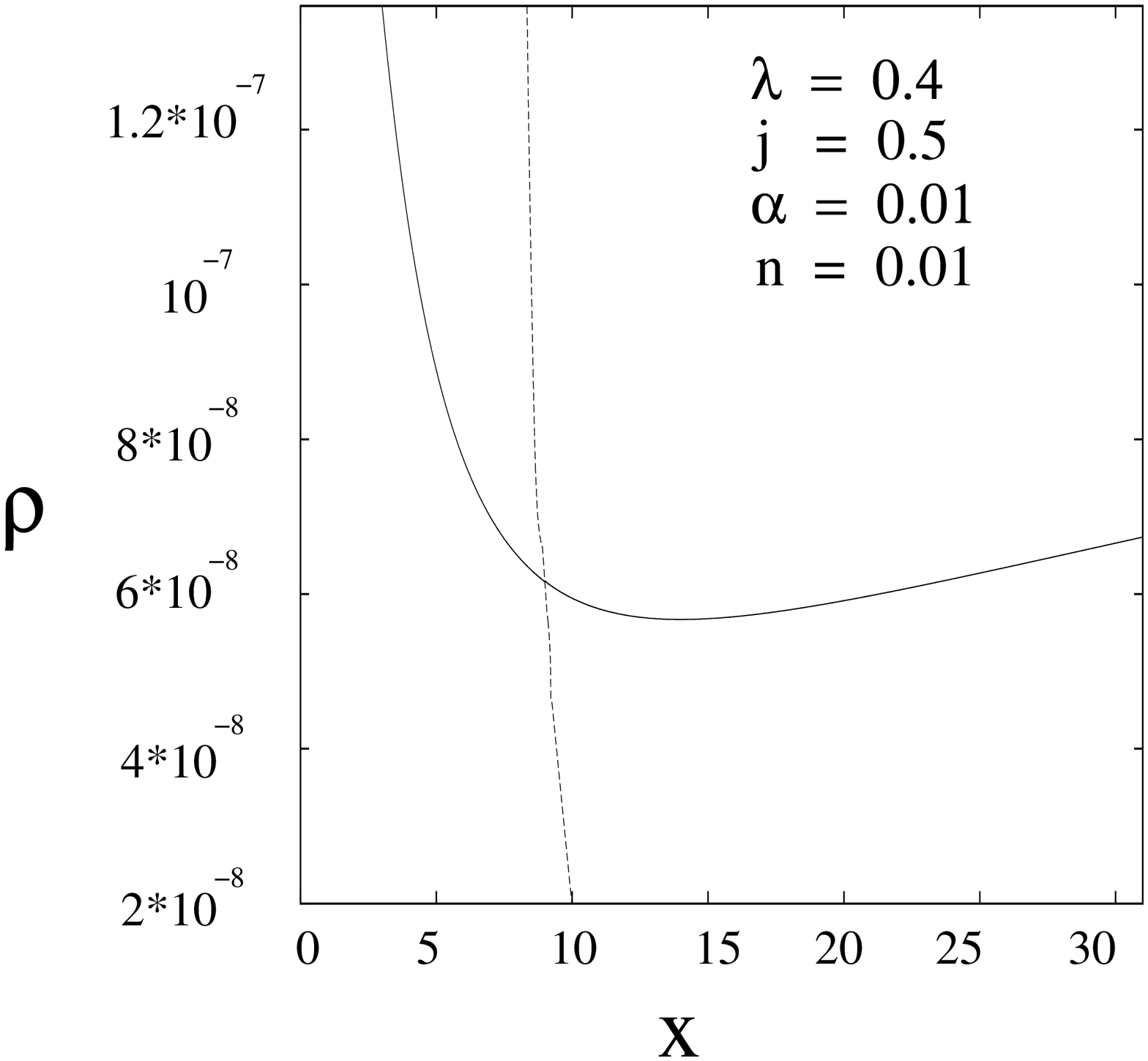}~~\includegraphics[height=1.6in, width=1.6in]{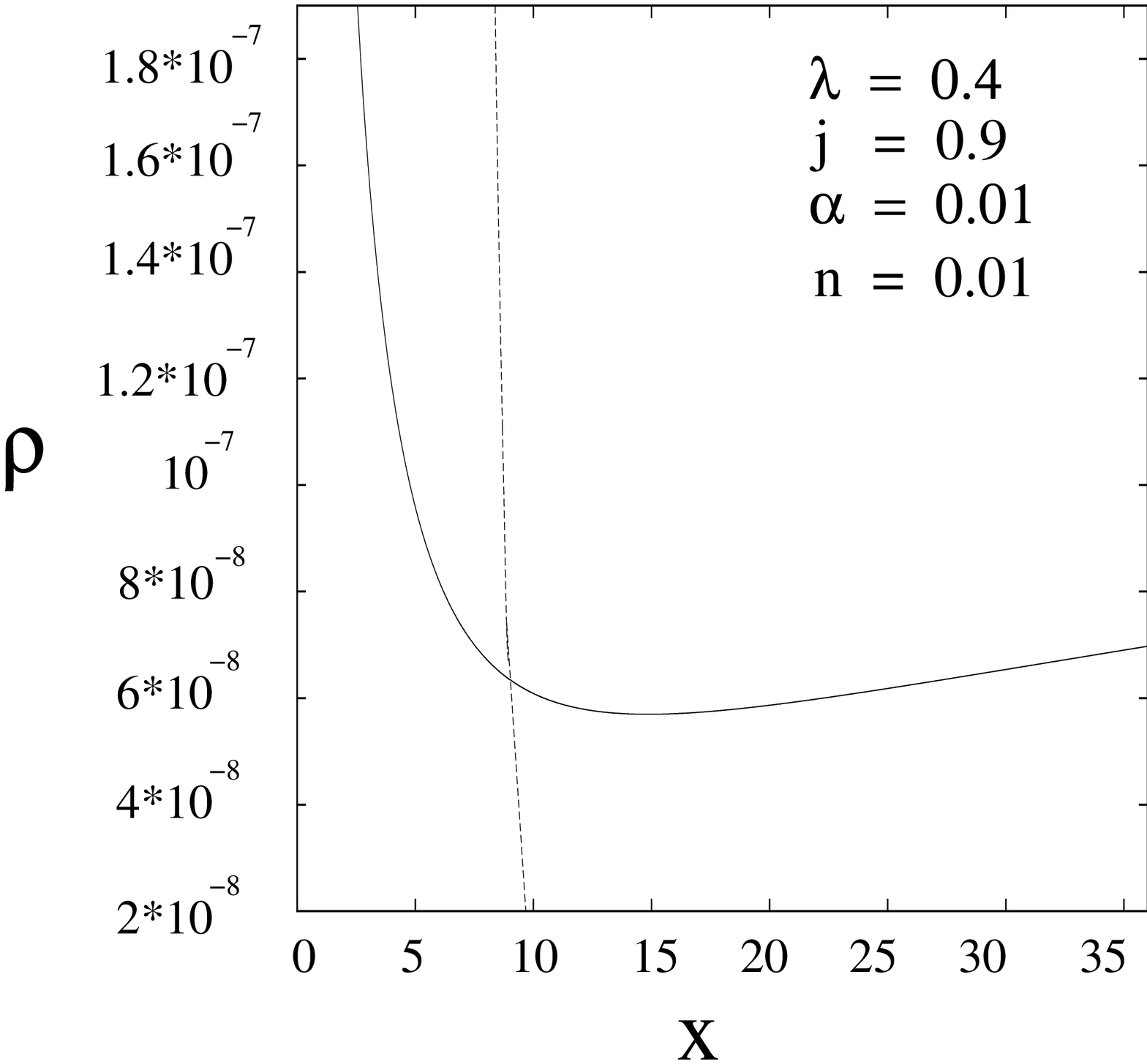}\\
\hspace{1cm}
Fig 2(a)-(l) represent the variation of the fluid density in accretion and wind flows as functions of radial coordinate corresponding to the velocity profiles shown in  (a) Fig 1(a), (b) Fig 1(b), (c) Fig 1(c), (d) Fig 1(d), (e) Fig 1(e), (f) Fig 1(f), (g) Fig 1(g), (h) Fig 1(hc), (i) Fig 1(i), (j) Fig 1(j), (k) Fig 1(k) and (l) Fig 1(l). The solid lines represent the accretion whereas the dotted lines are for wind.

\end{figure}

Now we have used velocity profiles \cite{Biswas1} of fig 1(a) to 1(l) and plotted $\rho$ vs x. These are given in 2(a) to 2(l).

If we look at fig 2(a) at first where we have drawn the density profile for spherical accretion of fluid of adiabatic nature upon Schwarzschild black hole. Here the fluid density along accretion branch slowly increases and is maximum at black hole event horizon . But in wind branch density increases very slowly when radial distance $x>x_{c}$, the sonic point distance. But when $x<x_{c}$ the density increases abruptly in the order of . $10^{6}-10^{7}$. In 2(b) the spherical accretion of modified Chaplygin gas upon Schwarzschild black hole has been studied. Here the density in wind branch is steeply increasing as we go near to the black hole. But the density in accretion branch is bit interesting . At first the density is decreasing as we go towards the black hole in the $x<x_{c}$ region. Then after reaching a local minima again the density increases but the rate of increasing is much lower than that of the wind branch. While to analyse 2(a) we can tell that the strong gravitational attraction of the black hole makes the fluid more denser as it accretes towards the black hole. It is to be noted that the density of wind fluid is much larger than that of accreting fluid. As at event horizon the fluid which are to be wind out has the velocity almost equal to zero as the strong gravitational force does not allow it to move out. So there the fluid is only gathered causing the increment of its density. Besides accreting fluid is been absorbed by the black hole which doesn't let the density to be increased.

On the other hand the case where modified Chaplygin gas is the accreting fluid under the same conditions as before we can see the slope of winding fluid is much much higher. We can explain it as we go outwards we get a negative pressure so as we go outwards we get a negative prewssure which decreases the value of density rapidly. To explain the accreting fluid density we can say that the accreting velocity at very far is very low as the negative pressure created by modified Chaplygin gas does not allow it to move inwards. So as the fluid is not moving inward or outwardhas a high density there. But as the fluid starts to accrete where gravitational pull is stronger than the negative force we will have a lower density but when again we are near to black hole again the density increases.

2(c), 2(e) and  2(g) has almost the same nature as 2(a)(2(c), 2(e) and 2(g) represent disc flow for j=0, 0.5 and 0.9 respectively). Besides the main features of 1(d), 1(f) and 1(h)(disc accretion of modified Chaplygin gas for j=0, 0.5 and 0.9) are quite similar with 2(b). The only difference to be noted that when we go far from the black hole the density is gradually lower for $j=0~ case>j=0.5~ case>j=0.9~ case$. This is very obvious again as increasing black hole spin creates increasing gravitational force which again increases the velocity of accretion which does not allow the fluid to gather at a place and force to move and the density decreases. Finally, it is to be noted carefully that the modified Chaplygin gas fluid density in 2(b),(d),(f),(h) are lower than the corresponding adiabatic cases 2(a),(c),(e) and (g) respectively. This is strongly supported by the negative pressure creating nature of modified Chaplygin gas.

At last we have reached to 2(i) to (l) which are $\rho$cs x graph corresponding to 1(i) to 1(l), i.e., the best fit values of $\alpha$ and $\beta$ which are supported by observational data. They all have the basic features as 2(b).

Dark energy accretion weakens the accretion and strengthens  the wind such that the accretion disc may become more and more weak : This was the conclusion of \cite{Biswas1}. In this paper where I am analysing the density profiles I can always see that the density along the wind branch is much much higher in Chaplygin gas case rather than the adiabatic case. This result resembles with the idea from \cite{Biswas1}. But the remarkable information in this case is in the accretion curve. Where, accretion branch has a local minima showing a minimal value of density near  the black hole. It is to be noted thatthis minima is not at the critical point. Particularly, I think this local minima signifies the CENBOL. From denominator of (\ref{7}) it is clear that from $D=0$ I do not have $u_{c}=c_{sc}$, rather $u_{c}$ also depends on different parameters like $\alpha$ and $n$ etc. According to the literature, beyond CENBOL matter again flow with the supersonic speed. For adiabatic cases (1(a), (c), (e) and (g)) it is very clear that identifying the radius where the CENBOL is not prominent. But for Chaplygin gas accretion case I can easily identify the radius (let $x_{CENBOL}$) where the CENBOL actually takes place and particularly I can particularly say that $x_{CENBOL}>x_{c}$. 
{\bf Acknowledge :}\\\\
I want to thank State Govt. of  West Bengal, India for awarding JRF, Mr. Aditya Kumar Naskar for different technical supports, prof. Subenoy Chakraborty, Dr. Ujjal Debnath and Ms. Nairwita Mazumder for different helps and Prof. Banibrata Mukhopadhyay for the idea. 

\end{document}